\documentclass[a4paper,11pt]{article}
\usepackage{geometry}
\renewcommand{\baselinestretch}{1.4} 

\usepackage{here}

\usepackage{multicol}
\usepackage{multirow}
\usepackage{fancybox, ascmac}

\usepackage[authoryear]{natbib}
\usepackage{here}
\usepackage[hyphens]{url}
\usepackage{hhline}
\usepackage{amsmath, amssymb} %数式用フォントはコレ
\usepackage{bm} %ギリシャ文字等も太字にできる\bm
\usepackage{array, booktabs} %表組用, 表の罫線用
\usepackage[utf8]{inputenc}
\usepackage{txfonts}
\usepackage[T1]{fontenc}
\usepackage{palatino}
\usepackage[dvipdfmx]{graphicx, color}
\usepackage{tikz}
\usepackage[framemethod=tikz]{mdframed}
\usepackage{hyperref}
\hypersetup{
    colorlinks=true,
    linkcolor=blue,
    citecolor=blue,
    filecolor=magenta,      
    urlcolor=cyan,
    pdfpagemode=FullScreen
    }

\usetikzlibrary{intersections,calc} % 垂線を描く際に必要
\usetikzlibrary{decorations.pathreplacing} % braceを描くために必要

\usepackage{tabularx} %表の枠の横幅指定\begin{tabularx}{幅}{列指定}
\usepackage{colortbl} %表の枠に色を付ける
\usepackage{float} %図表の配置で必要

\def\diamondleaders{\par\vskip.5\baselineskip
 \leavevmode\hbox{}\hskip1zw\leaders
 \hbox to.5zw{\hss\footnotesize ◇\hss}
 \hfill\hskip1zw\hbox{}\par}
\makeatletter
\def\thanks#1{%
   \footnotemark
   \edef\@tempa{\noexpand\noexpand\noexpand\footnotetext[\the\c@footnote]}%
   \toks@\expandafter{\@thanks}%
   \toks\tw@{{#1}}
   \xdef\@thanks{\the\toks@\@tempa\the\toks\tw@}}
\makeatother

\begin{document}
%\title{Sustainability of Social Security for Future Japan:\\
%What is the key to resolve demographic puzzle?}

\title{The Impact of the Social Security Reforms on Welfare: \\
Who Benefits and Who Loses Across Generations, Gender, and Employment Type?}
%\title{What are the Social Security Reforms \\
%that Mitigate Welfare Losses in Japan?}
\author{{\large Hirokuni Iiboshi% (Nihon University)
\thanks{Nihon University, College of Economics. 
E-mail: \url{iiboshi.hirokuni@nihon-u.ac.jp}}\quad \& \quad 
Daisuke Ozaki% (Tokyo Metropolitan University)
\thanks{Tokyo Metropolitan University, Graduate School of Management.
E-mail: \url{ozaki-daisuke@ed.tmu.ac.jp}}
\thanks{We specially thank Sagiri Kitao for her valuable comments as a discussant.
We would like to thank Masataka Eguchi, Ryo Hasumi, Kazuki Hiraga, 
Yasuo Hirose, Munechika Katayama, Hiroyuki Kubota, Yoshiyuki Nakazono, 
Kenji Miyazaki, Toshiaki Ogawa, Tetsuaki Takano,
Kazuhiro Teramoto, Takayuki Tsuruga, Hikaru Saijo, Mototsugu Shintani, Naoto Soma, Kozo Ueda and seminar participants at Hosei University 2021, Osaka University 2022, Japan Economic Association Fall Meeting 2022 for their useful comments and suggestions. 
Iiboshi is grateful for financial support from the Japan Society for the Promotion of Science (21K01464). All remaining errors are our own. }}}

\maketitle

\vspace{1\baselineskip}

\begin{abstract} %要修正
\begin{small}
\noindent
%我々は、性別と雇用形態で区別した4タイプのエージェントが存在する世代重複モデルを用いて、急速な高齢化と世界で最も高い政府債務残高に直面する日本で社会保障改革が実施された場合の影響を定量的に分析する。
%その結果、定年延長を伴わない社会保障改革は将来世代の厚生を引き上げる一方で、特に医療・介護支出の自己負担引き上げは現役世代の低所得層（女性、パートタイマー）の厚生を顕著に引き下げることを見出した。
%一方、年金代替率引き下げの改革ではフルタイマーの厚生低下がより大きくなっていた。
%社会保障改革と定年延長の一体化により、現役世代の厚生が現行よりも2〜5%程度まで改善することが期待できる。
Japan is facing rapid aging and the highest government debt among developed countries.
We quantitatively explore the impact of social security reforms in Japan using an overlapping generations model with four types of households distinguished by gender and employment type.
The results of our paper suggest that reducing social security benefits without extending the retirement age raises future generations' welfare while lowering the current generation's welfare. 
In particular, a medical and Long-term care insurance reform significantly lowers the welfare of current low-income groups: females and contingent workers.
In contrast, a reform reducing the pension replacement rate leads to a more significant decline in the welfare of regular workers than contingent workers.
Combining social security reforms with an extension of the retirement age would mitigate the declining welfare of the current generation by about 3 to 4\%.

\vspace{0.5\baselineskip}

\vspace{0.5\baselineskip}
\end{small}
\end{abstract}
\newpage

%%%目次＝確認用
%\tableofcontents
%\newpage

%%本文スタート（本文も分けずにここに）%%%%%%%%%%%%%%%%%%%%%%%5

\section{Introduction}
\label{sec:INTRO}

%\subsection{Demographic transition in Japan}

%多くの主要先進諸国は人口の高齢化に直面している。その中でも、日本は高齢化の進行が顕著に早い。
%2018年時点で、日本の老年人口比率（15～64歳人口に占める65歳以上人口の割合）は47.2\%である。
%加えて、国連による2065年時点の同比率の推計値は、日本では約75\%と、近年急速に高齢化が進む韓国では80%強と、主要先進国の中で最も高い水準となる見通し（projection）が示されている（中位推計）。
%\footnote{総務省統計局「人口推計（2018年10月1日）」、United Nations, “World population prospects 2019”.}
%This paper focuses on the impact of social security reforms on social welfare in aging and a declining population in Japan.
Many developed countries are facing population aging. 
In particular, Japan is aging significantly faster than the rest of the world.
In 2018, Japan's elderly population ratio (the ratio of the population aged 65 and over to the population aged 15--64) was 47.2\%.
In addition, the United Nations' projection of the ratio in 2065 shows that it is expected to be the highest among developed countries (medium projection), at around 75\% in Japan and more than 80\% in South Korea, where the population has been aging rapidly in recent years.
\footnote{
Ministry of Internal Affairs and Communications, Statistics Bureau, "Population Estimates (October 1, 2018)"; United Nations, "World Population Prospects 2019".}

\begin{figure}[tb]
 \centering
 
 \includegraphics[width=80mm]{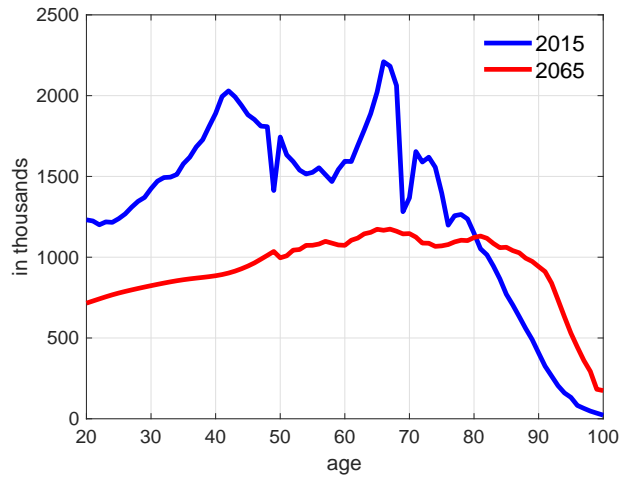}
 \vspace{-0.5\baselineskip}
 \caption{Change in the population distribution in Japan: from 2015 to 2065}
 \label{fig:pop_ds}
\end{figure}

%高齢化の進行は、将来の人口分布を大きく変化させる。
%図\ref{fig:pop_ds}は、日本における2015年の人口分布と、国立社会保障・人口問題研究所の予測に基づく2065年の人口分布を示している。
%2015年の人口分布には、2つのピークがある（青の実線）。
%これらを構成するのは、「団塊の世代（the baby boomer generation)」と、彼らの子の世代に当たる「団塊ジュニア世代 (children of the baby boomer generation)」である。
%2020年時点では、2015年に65歳に到達し始めた団塊の世代の多くは、すでに公的年金受給者となっている一方、団塊ジュニア世代は労働力の担い手である。
%しかし、2040年頃には、後者も65歳を超えて年金受給開始年齢に到達する。
%その結果、2030年台後半から2070年台にかけて、年金支給総額が大きく上昇する可能性が、政府の見通しでも指摘されている。
%\footnote{厚生労働省「2019年 将来の公的年金の財政見通し（財政検証）」（https://www.mhlw.go.jp/stf/seisakunitsuite/bunya/nenkin/nenkin/zaisei-kensyo/index.html　2021年3月2日アクセス）。}
%加えて、図\ref{fig:pop_ds}は高齢化の進行だけでなく人口の減少、つまり労働力の減少も示唆している。
Population aging significantly changes the future distribution of the population.
Figure \ref{fig:pop_ds} shows the population distribution in Japan in 2015 and in 2065 based on the projections from the National Institute of Population and Social Security Research (IPSS).
The population distribution in 2015 has two peaks (blue line).
These are the baby boomer generation and the children of the baby boomer generation.
As of 2020, many of the baby boomers who began reaching the age of 65 in 2015 will already be public pensioners.
Meanwhile, the junior baby boomers are the labor force.
However, around 2040, the latter will also reach the retirement age of 65 and begin receiving public pensions.
As a result, the total amount of pension payments could rise significantly from the late 2030s to the 2070s, according to government projections.
\footnote{
Ministry of Health, Labour and Welfare, "2019 Financial Projection of the Public Pension (Fiscal Verification)" 
(\url{https://www.mhlw.go.jp/stf/seisakunitsuite/bunya/nenkin/nenkin/zaisei-kensyo/index.html}\quad accessed March 2, 2021).}

\begin{figure}[H]
 \centering
 \noindent\begin{minipage}[t]{1\columnwidth}%
 \begin{center}

\hspace{2em}
\includegraphics[width=65mm, height=62mm]{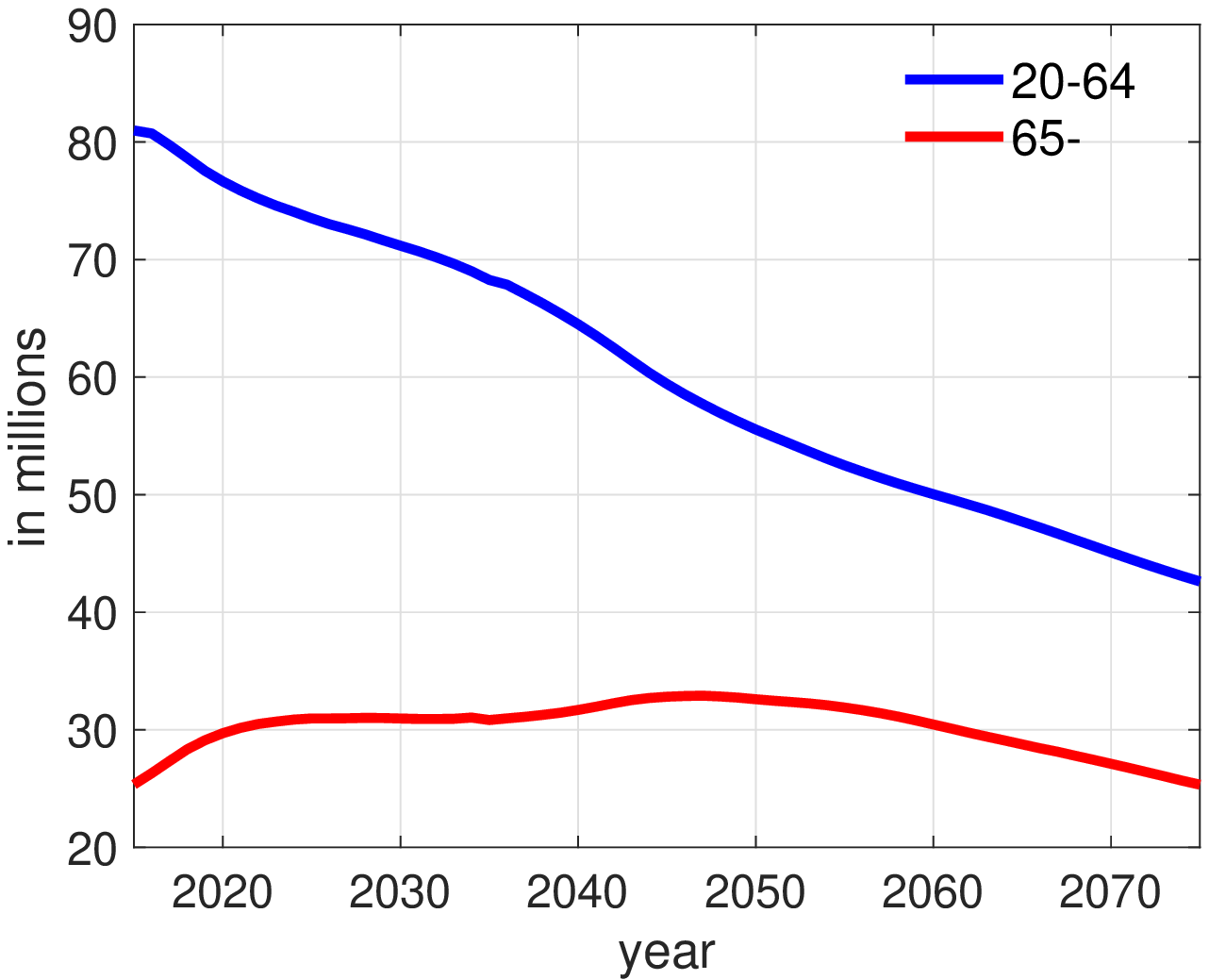}~~~~~
\includegraphics[width=65mm, height=62mm]{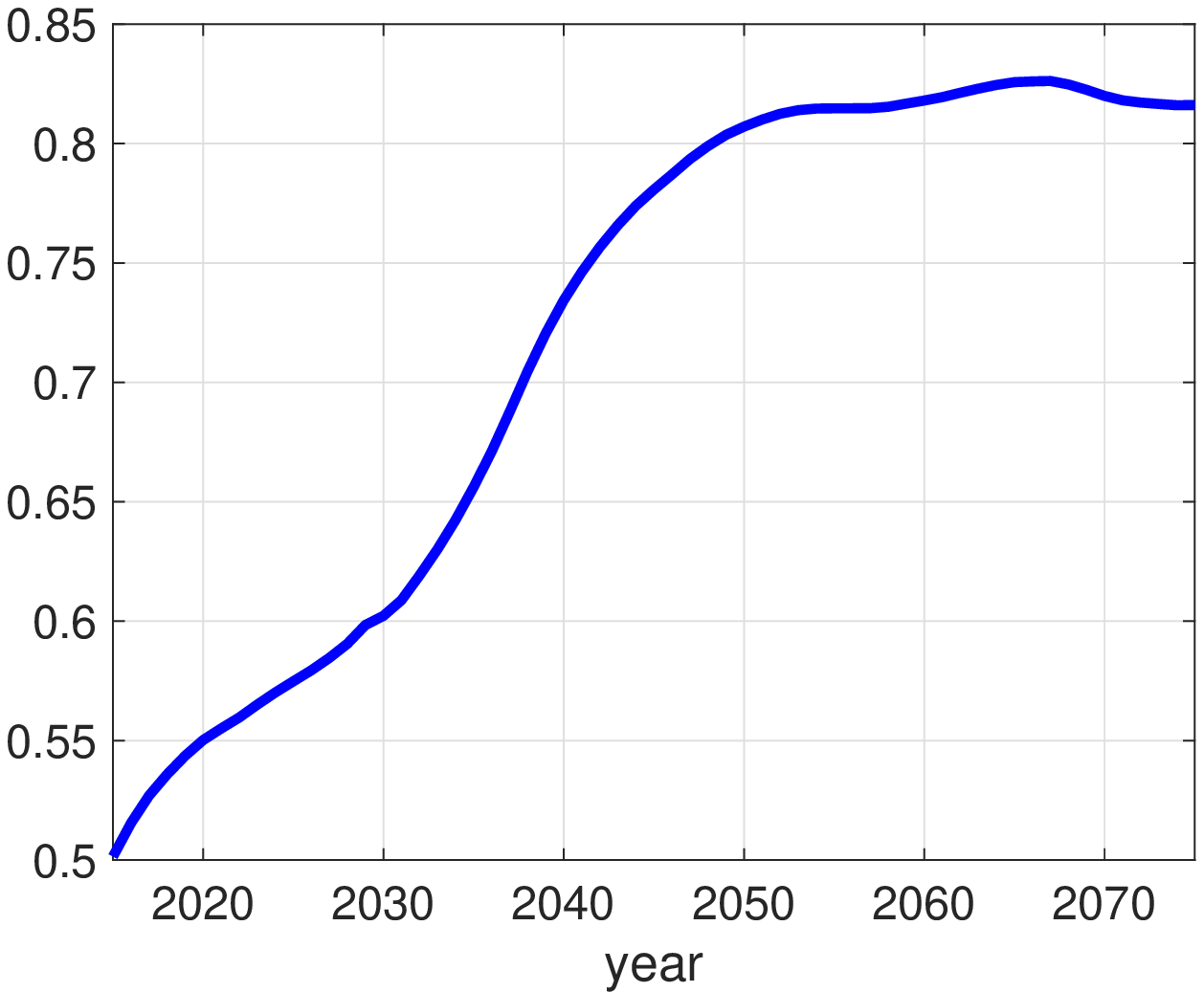}

(a)~Demographic change \hspace{7em}
(b)~Dependent rate ~~~
\end{center}
\end{minipage}

 \caption{Demographic transition in Japan: 2015--2075}
 \label{fig:DTJ}
\end{figure}

%加えて、高齢化の進展に伴う人口減少は労働力の減少を引き起こす。
%国立社会保障・人口問題研究所（National Institute of Population and Social Security Research）のprojectionでは、主に勤労者が占める20～64歳人口は将来に向けて単調に減少していく一方、主に引退者が占める65歳以上人口には増加傾向にある。（図\ref{fig:DTJ}左パネル）。
%この傾向は、従属老年人口比率（old-age dependency ratio）の推移に顕著に反映されている。
%\footnote{ここで、従属老年人口比率$\equiv$ 65歳以上人口/20～64歳人口。}
%この比率は2015年では50%ほどであるが、2050年頃には80%を超える見通しが示されている（図\ref{fig:DTJ}右パネル）。
%こうした人口構造の変化は、公的年金や医療・介護保険制度の運営に大きな影響を及ぼすため、日本では重要な政策課題とされてきた。そして今日まで、高齢化の影響を定量的に分析する研究が、数多く蓄積されてきた。
In addition, the population decline associated with graying leads to a decrease in the labor force.
According to the projection by the IPSS, the population aged 20--64, consisting mainly of workers, will decrease monotonically in the future, while the population aged 65 and over, consisting mainly of retirees, will increase, as shown in Figure \ref{fig:DTJ} Panel (a).
This trend is reflected clearly in the transition of the old-age dependency ratio.
\footnote{ 
The old-age dependency ratio is defined as the ratio $\equiv$ population aged over 65 $/$ population aged 20--64.}
While this ratio was about 50\% in 2015, it is projected to exceed 80\% by around 2050, shown in Figure \ref{fig:DTJ} Panel (b).
Because these demographic changes will exert significant impacts on the public pension and the medical and LTC insurance systems, this topic must be considered as one of the most urgent challenges on the policy agenda in terms of fiscal sustainability. 

%%%%%%%%%%%%%%%%%%%%%%%%%%%%%%%%%%%%%%%%%%%%
%%%%%%%%%%%%%%%%%%%%%%%%%%%%%%%%%%%%%%%%%%%%% 

%%%%%%%%%%%%%%%%%%%%%%%%%%%%%%%%%%%%%%%%%%%%
%%%%%%%%%%%%%%%%%%%%%%%%%%%%%%%%%%%%%%%%%%%%% 

%こうした事実や見通しに基づき、日本政府は賦課方式（pay-as-you-go: PAYG）的な制度に基づいて運営されている社会保障、特に公的年金、医療保険、介護保険の財政を維持可能なものとするために、さまざまに改革を実施してきた。
%たとえば、公的年金の支給開始年齢の引き上げ、給付水準を段階的に調整するマクロ経済スライド（“macroeconomic slide”）の導入、また、医療費適正化計画（the plan for appropriate medical care expenditures）に基づく病床整理等を通じた医療費の抑制策の推進などであり、現在もさらなる政策が議論されている。
%日本では、政府の財政健全化は長年にわたり重要な政策課題とされてきた。その理由は、多額の政府債務の存在にある。2000年代以降、日本の政府債務残高（対GDP比）は主要先進諸国の中でも最も高く、2018年時点で237.16\%となっている。
%\footnote{財務省「財政に関する資料」（https://www.mof.go.jp/tax_policy/summary/condition/a02.htm　2021年3月2日アクセス）。}
%ただしこうした社会保障改革は、今後は人口と労働力が減少し、産出（output）も減少する中で行われることになる。本稿ではこの点に着目し、人口減少下で実施される社会保障改革が人々の厚生にどのような影響を及ぼしうるかを、いくつかのシナリオを設定して定量的に分析する。
%According to IPSS, 
In response to the IPSS projection, the Japanese government has been reforming various social security systems, especially the public pension, medical insurance, and long-term care (LTC) insurance which run on a pay-as-you-go (PAYG). 
For example, the government has raised the retirement age for public pension benefits, introduced the "macroeconomic slide" to adjust the benefit levels gradually, and promoted measures to control medical costs by consolidating hospital beds based on the plan for appropriate medical care expenditures.
Because the government debt resulting from these %medical and LTC cost burdens 
costs of social security expenditures has accumulated enormously over the years, casting a dark shadow on fiscal sustainability, reconciling them is a challenge for fiscal policy.

Since the 2000s, the government debt to GDP ratio has been the highest among developed countries, standing at 237.16\% as of 2018.
\footnote {
Ministry of Finance, "Documents on Public Finance," 
(\url{https://www.mof.go.jp/tax_policy/summary/}\\ \url{condition/a02.htm} \quad  accessed March 2, 2021).}

%%%%%%%%%%%%%%%%%%%%%%%%%%%%%%%%
%\subsection{What we do}

%上記の課題に答えるために、我々は米国における高齢化と公的年金改革の影響を分析した■McGrattan and Prescott (2017)のモデルを参考に、上記の問題を分析する。
%そのモデルとは、資産階級によって4タイプに区別された家計が存在する一般均衡型の重複世代モデルである。そこでは、法人部門と個人事業主部門（household business sector）の2部門、有形資産と無形資産、累進的な所得課税が考慮されている。
%本論文では、このモデルを我々の問題意識にあわせて拡張する。
%我々はそのモデルを用いて、日本の社会保障制度に基づき、人口および労働力の減少が将来のoutput等のaggregate variablesに及ぼす影響を推計する。そして、年金・医療・介護制度の改革が人々の厚生に及ぼす影響を定量的に検証する。
%■McGrattan and Prescott (2017) モデルの特徴は、マクロ経済データとミクロの所得データを用いて、モデルのパラメターを詳細に設定できることである。たとえば、このモデルでは労働所得に対して現実に基づいた累進的な税率をする。
%そのため、このモデルを日本の文脈に応用することで、より現実的な仮定のもとで政策変更の影響を検証できる。

\paragraph{Purpose and contribution of our paper}
To tackle these challenges, we compare and evaluate changes in social welfare by selecting one of several options for social security reforms subject to the fiscal sustainability constraint, following \citet{MP2017QE}.
To measure the effects of aging and public pension reform in the U.S., they use a general equilibrium overlapping generations (OLG) model with four types of households distinguished by wealth class.
The feature of their model is rich parameterization, which enables calibration corresponding to both macro-level data and micro-level income data, for example the application of the progressive tax bracket to the labor income tax rate.
By applying their model to Japanese policy reforms, our study provides an empirical outcome in more practical situations than previous Japanese studies.

%我々の研究では、■McGrattan and Prescott (2017) のモデルを、家計を性別（男性、女性）と雇用形態（フルタイマー、パートタイマー）で区別された4タイプとし、公的年金だけでなく医療・介護保険もを含める形に拡張した。
%そして、我々は以下で述べる複数のシナリオに基づいて年金・医療・介護制度改革を実施した場合の政策シミュレーションを行い、人口減少と社会保障改革が現在の勤労世代と引退世代、および将来世代の家計の厚生に及ぼす影響を定量的に分析する。
%その際、特に4タイプに区別されたエージェントの特性ごとのインパクトの違いに着目する。
%一般的には、男性よりも女性が、フルタイマーよりもパートタイマーの方が、賃金水準が低く、社会保障給付の削減や税・保険料負担の引き上げに対してより脆弱であると予想される。
We also extend the model of \citet{MP2017QE} to four types of households distinguished by gender (male and female) and employment types (regular and contingent worker), as well as including medical and LTC insurance besides public pensions.
We then quantitatively analyze the impact of social security reforms associated with a decreasing population on the welfare of different generations: the current working, retired, and future generations. 

%要素8: 主な結果
%本論文では、次の2つのシナリオを設定する。すなわち、1つ目は現状の制度に基づくベースライン・シナリオ、2つ目は65歳から70歳への定年延長シナリオである。そして、これらのシナリオのもとで、次の3つの改革が行われた場合に基づいた政策シミュレーションを行う：
%(1) 年金所得代替率を現行の62%から2047年まで段階的に50.8%へ引き下げる改革、(2) 高齢者の医療費自己負担率を全年齢で一律3割に引き上げる改革、(3)介護費自己負担率を一律3割に引き上げる改革。
In our paper, reforms of the pension, medical insurance, and LTC insurance are considered according to two scenarios: (1) a baseline scenario based on the current system, 
(2) a scenario with the extension of the retirement age from 65 to 70.
Within these two scenarios, we simulate three options for the reform: 
(1) a reform to lower the pension income replacement rate gradually from the current 62\% to 50.8\% by 2047; 
(2) a reform to raise the copayment rate for medical expenses for the elderly uniformly to 30\% for all ages; 
and (3) a reform to increase the copayment rate for LTC expenses uniformly to 30\%.

%その結果、人口とoutputの水準が最も低下する2040年代半ばには、outputは2020年の75\%程度の水準にとなること、そして定年延長や社会保障改革がGDPを引き上げる効果は限定的であることが示された。
%一方で、定年延長は特に現在の勤労世代の厚生を改善した。社会保障改革に着目すると、年金所得代替率の引き下げはフルタイマーの厚生を相対的に大きく引き下げ、医療・介護の自己負担率引き上げは現在の引退世代の厚生を引き下げる結果となった。
%特に、医療・介護改革では特に所得・資産の低い女性パートタイマーの厚生を大きく引き下げることが示された。
According to our simulation, by the mid-2040s, when the population and output levels are at their lowest, the output will be about 75\% of its 2020 level, and extending the retirement age and reforming social security systems will have little effect on increasing the GDP.
However, extending the retirement age will improve the welfare of the current generations in particular.
Focusing on social security reforms, the reduction in the pension income replacement rate will reduce the welfare of regular workers by a relatively large amount, and the increase in the copayment rate for medical and LTC expenditures will lower the welfare of the current retired generation.
We also find that medical and LTC insurance reform will significantly lower the welfare of females and contingent workers with relatively low incomes and assets.

% 要素6: 研究の貢献
%本研究の貢献の1つは、財政維持可能性のために不可欠とされる社会保障改革を実施する際に、どのようなタイプの家計に配慮すべきかを定量分析に基づいて示している点である。
%加えて、2つのシナリオと3つの社会保障改革が人々の厚生に及ぼす影響を、現在の勤労・引退世代、および将来世代における4タイプのエージェントに区別して検証することで、より現実的な政策含意を提供している。
The contributions of our study are as follows.
The first is to quantify which characteristics of households are more significantly negatively affected in reforming the social security system needed for fiscal sustainability.
The second is to simulate the impact of the two scenarios and the three social security reforms on household welfare, distinguishing between generations and types of households.
Our simulations allow us to provide more practical policy implications.

% 要素10: 本文構成の紹介
%本論文の残りは次のように構成される。
%第2節では日本の人口動態に関する事実を整理したうえで、先行研究に対する本研究の貢献を示す。
%第3節では我々が構築したモデルについて述べる。
%第4節では本分析で考慮するデータと制度的背景に基づいてカリブレーションとパラメターの設定について説明し、定常状態を計算する。
%第5節では定常状態の結果に基づいて移行過程を計算する。そこで、上述した複数のシナリオと社会保障改革を想定したシミュレーションを行う。
%第6節では本分析のconclusionと今後の課題を示す。
The rest of this paper is organized as follows.
Section \ref{sec:LITERA} summarizes the institutional background of Japanese social security system and shows the contribution of this study to the previous research.
Section \ref{sec:MODEL} describes our model.
Section \ref{sec:CALIB} explains the calibration and parameterization based on the data and institutional background for the analysis, and calculates the steady state.
Section \ref{sec:NA} computes the transition path based on the steady state. Then, we conduct simulations setting the various scenarios and social security reforms described above.
Section \ref{sec:CONCL} discusses the conclusions of our paper and suggests future research.

%%%%%%%%%%%%%%%%%%%%%%%%%%%%%%%%%%%%%%%%%%%%%%%%%%%%%%%%%%%%%%%%%%%%%%%%%%%%%%%%%%%%
\section{Background and related literature}
\label{sec:LITERA}

%本節では、次の2つの点について記述する。
%第1に、日本のdemographic factsおよび1人当たりの医療費と介護費用を示して、本論文が対象とする課題を明確にする。
%第2に、関連する先行研究を整理したうえで、それらの中での本論文の貢献を示す。
In this section, we describe two factors that provide the motivation and background for our study.
First, we present the demographic data and medical and LTC expenditures per capita to clarify the arguments on which our paper focuses.
Second, we summarize the literature review for the relevant field, in particular for Japan.

%%%%%%%%%%%%%%%%%%%%%%%%%%%%%%%%
\subsection{Institutional Background}

%日本では、国民は何らかの公的医療保険に加入することが義務付けられている。
%医療費の自己負担率は、2021年では、原則として、70歳未満の者は3割、70歳以上75歳未満は2割負担、75歳以上は1割負担と定められている（なお、6歳未満の未就学児は2割負担）。
%ただし2021年6月4日に改正法の成立により、医療保険における被保険者の負担軽減を意図して、2022年後半から75歳以上の高齢者の自己負担を2割に引き上げられることが決まった。
%また、介護保険は、40歳からの加入が義務付けられている。
%介護費用の自己負担率は40歳以上の全年齢で1割負担とされている（図\ref{fig:Med-LTC_copay}）。

\begin{figure}[tp]
 \centering

 \includegraphics[width=110mm]{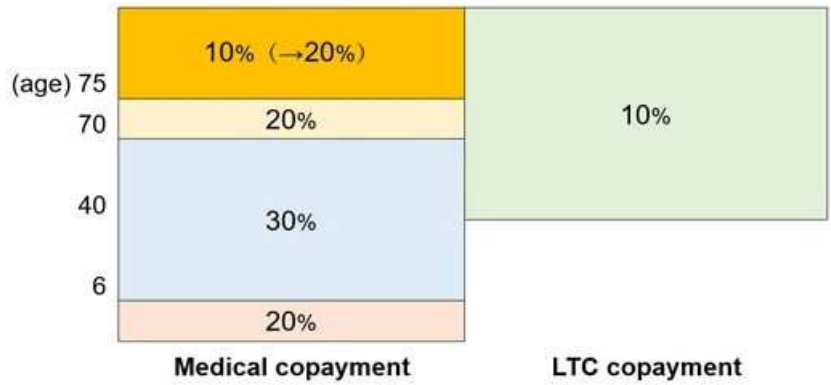}

 \vspace{-0.3\baselineskip}
 \caption{Medical and LTC copayment rate}
 \label{fig:Med-LTC_copay}
\end{figure}

In Japan, citizens are required to be covered by some form of public medical insurance.
As of 2021, the copayment rate for medical expenses is set at 30\% for those under 70 years old, 20\% for those between 70 and 75 years old, and 10\% for those over 75 years old (preschool children under 6 years old are required to pay 20\%).
However, due to the passing of a revised law on June 4, 2021, it was agreed that the copayment for those aged 75 and over will be rised to 20\% in the latter half of 2022, to reduce the burden on the covered persons in the medical insurance system.
In addition, LTC insurance is mandatory from the age of 40.
The copayment rate for LTC expenditures is set at 10\% for everyone aged 40 and over (Figure \ref{fig:Med-LTC_copay}).

%%%%%%%%%%%%%%%%%%%%%%%%%%%%%%%%%%%%%%%%%%%%
%%%%%%%%%%%%%%%%%%%%%%%%%%%%%%%%%%%%%%%%%%%%% 

%日本では2000年に公的介護保険が導入された。
%その後、高齢化とともに介護保険財政の維持可能性への懸念が強まり、現在までにたびたび制度変更が実施された。
%公的介護保険導入以降、介護サービスの利用者は3倍強増加し、要介護度別認定者数は2000年の218万人から2018年には658万人へと、約3倍に増加した。
%\footnote{厚生労働省「介護保険制度の概要」（https://www.mhlw.go.jp/stf/seisakunitsuite/bunya/hukushi_kaigo/kaigo_koureisha/gaiyo/index.html　2021年3月2日アクセス）。}
%介護費用は増加傾向にある。そのため、次のような介護保険給付の抑制改革が実施されてきた（■Fu et al, 2017）。すなわち、認定された要介護度に応じた一部サービスの利用制限、現役世代と同程度の所得を得ている高齢者の自己負担率の引き上げ、などである。
%% https://econpapers.repec.org/article/eeejhecon/v_3a56_3ay_3a2017_3ai_3ac_3ap_3a103-112.htm
Public LTC insurance was introduced in Japan in 2000.
Afterward, there were increasing concerns about the financial sustainability of LTC insurance with the aging of the population, and the program has been reformed frequently since then.
After introducing public LTC insurance, the number of users of LTC services, certified by the level of care required, more than tripled from 2.18 million in 2000 to 6.58 million in 2018.
\footnote{
Ministry of Health, Labour and Welfare, "Summary of the LTC Insurance System" (\url{https://www.mhlw.go.jp/stf/seisakunitsuite/bunya/hukushi_kaigo/kaigo_koureisha/gaiyo/}\\ \url{index.html} \quad accessed March 2, 2021).}
Correspondingly, the cost of LTC is on the rise, so the government is trying to implement reforms reducing the LTC insurance benefits, for example limiting the utilization of services according to the level of LTC required, increasing in the copayment rate for the elderly people earning the equivalent level of income as the working-age generations, and so on.
\footnote{
\citet{Fuetal2017} analyze the impact of these LTC insurance reforms on labor supply using micro-level data.}

%上記のような、将来の財政維持のため社会保障改革は、家計のライフサイクルを通じた消費や貯蓄に大きな影響を及ぼす可能性がある。
%さらにその影響は、資本蓄積や産出（output）などのaggregate variablesにも反映される。
%加えて、人口や労働力の減少は経済成長の負の影響を及ぼす可能性がある。
%実際、内閣府「中長期の経済財政に関する試算」や日本銀行「経済・物価情勢の展望」など、政府が発表する試算は近年のGDP成長率の鈍化や、TFP成長率の低下が示している。
%たとえば、日本銀行の試算では2015年以降は0.01～0.43%と特に低い値が示されている。
%\footnote{日本銀行「需給ギャップと潜在成長率」\url{https://www.boj.or.jp/research/research_data/gap/index.htm/} (2021年6月29日アクセス)。}
%そのため、さらなる社会保障改革は人口減少と経済成長のさらなる鈍化が同時に発生する時期に行われることになると考えられる。
%そして、我々はそうした改革が人々の消費・貯蓄行動や厚生に及ぼす影響を検証することは、今後の政策議論にも重要な示唆をもたらすと考える。
%% ★コロナについてはコメントしてエスケープ。
%\footnote{もちろん、COVID-19は社会保障の維持可能性や人口動態に大きな影響を及ぼす。しかし、我々はより長期的な視点での人口動態に着目して社会保障改革の影響を分析することを目的としている。そのため本研究は、COVID-19の影響は考慮せず、2019年までのデータを用いることとした。}
The social security reforms to maintain fiscal sustainability could have a significant impact on aggregate flows such as consumption and savings over the household life cycle.
The reforms would also affect aggregate stock variables such as capital accumulation, which form future output.
Since a decreasing population and labor force could also hurt economic growth, it is necessary to examine the effect of reforms comprehensively.
In fact, the outlooks isssued by the government, such as the Cabinet Office's "Medium- and Long-Term Economic and Fiscal Outlook" and the Bank of Japan's "Outlook for Economic and Price Conditions," indicate a slowdown in the GDP growth in recent years and a decline in the total factor productivity (TFP) growth rate.
For example, the Bank of Japan's outlook shows particularly low values ranging from 0.01--0.43\% from 2015 onward.
\footnote{
Bank of Japan, "Supply-Demand Gap and Potential Growth Rate" 
(\url{https://www.boj.or.jp/research/}\\ \url{research_data/gap/index.htm/} \quad accessed June 29, 2021).}
Therefore, additional social security reforms are likely to be implemented simultaneously with a decreasing population and a further slowing of economic growth.
It is evident that measuring the effects of such reforms on people's welfare, evaluated from the amount of consumption and saving behavior, could have important implications for future policy discussions.
\footnote{
Of course, COVID-19 will significantly affect the social security systems and demographic dynamics.
However, we aim to analyze the impacts of social security reform by focusing on demographics from a longer-term perspective. For this reason, our study does NOT consider the effects of COVID-19 and uses data prior to 2019.} 
%% ★コロナについては上の注でコメントするのみ。
%% 以下、簡単に無形資産＋Mcgrattan (2020RED)追加
On the other hand, although the growth rates of TFP and GDP are forecasted to be lower, there is a perception that this is due to the underestimation of the magnitude of GDP. 
According to \citet{mcgrattanIntangible2020a},  intangible assets, not included in GDP,  and multi-sector production significantly contribute to economic fluctuations.  
As we will see in more detail later, \citet{MP2017QE} analyzed social security reform in an aging economy by incorporating intangibles and multiple sectors into a general equilibrium OLG model. And we follow their model and approach.

\subsection{Fiscal sustainability and the declining labor force}
%高齢化が財政運営に及ぼす影響を分析した論文は数多く蓄積されてきた。
%たとえば、日本の高齢化と財政の維持可能性を扱った研究として、■Hansen and Imrohoroglu (2016) がある。
%彼らは新古典派成長モデルを用いた分析を行った。
%そして彼らは、財政を維持するために、課税対象（tax base）の拡大や社会保障給付抑制等の改革を実施したときは40\%程度の消費税率が必要であることを見出した。
%さらに彼らは、それらの改革を実施しなかった場合は、60\%程度もの消費税率が必要になるという結果も示した。
There is a large amount of literature analyzing the impact of population aging on fiscal administration in Japan.
\citet{HanIm2016RED} analyze aging and fiscal sustainability using a neoclassical growth model. 
They find that a consumption tax rate of about 40\% is needed to sustain the fiscal balance when reforms such as the expansion of the tax base and the reduction of social security benefits are implemented. 
They also show that a consumption tax rate of about 60\% would be needed if those reforms are not introduced.

%医療・介護をモデルに組み入れた研究も数多く存在する。
%たとえば、日本に焦点を当てた■Braun and Joines (2015) は、一般均衡型の世代重複モデルを用いて、高齢化が公的年金、医療保険、介護保険を含む日本の社会保障運営に与える財政的なインパクトを推計した。
%そして彼らは、消費税の引き上げや医療支出の自己負担率の引き上げなどの改革が財政維持可能性（fiscal sustainability）に与える影響を分析した。
%彼らは、財政を消費税率でバランスさせるという設定で政策シミュレーションを行い、財政を維持するためには30～45\%程度の消費税率が必要となることを示した。
%さらに、高齢者の医療の自己負担率を30\%に引き上げる改革を同時に行えば、財政維持に必要な消費税率は20\%程度にとどまることを見出した。
%There are a lot of studies that incorporate medical and LTC into the model. % remove 2022/04/25

\citet{BJ2015JEDC} investigate the fiscal impact of an aging population on Japan's social security system, including public pensions, medical insurance, and LTC insurance using a general equilibrium OLG model.
They also analyze the impact on fiscal sustainability of reforms such as raising the consumption tax and increasing the copayment rate for medical expenditures.
They conduct policy simulations under the assumption that public finances are balanced by the consumption tax rate, and they show that consumption tax rates of 30--45\% would be necessary to sustain the fiscal balance.
Furthermore, they find that, if reforms to raise the copayment rate for medical expenditures for the elderly to 30\% are to be implemented simultaneously, the consumption tax rate needed to sustain the fiscal balance would be around 20\%.

%■Kitao (2015 JEDC) は年金制度改革に着目しつつ，日本のdemographic changeが財政に与える影響を定量的に分析した。
%この研究は、年金改革を行わない場合は財政維持に必要な消費税率が最大で50\%近くまで上昇する一方で、年金支給開始年齢の引き上げと20\%の給付削減を併せて実施すれば、消費税率の上昇は30\%弱にとどまることを示した。
%加えて、■Kitao (2018) は年金制度改革にフォーカスし、年金給付を削減する改革を始めるタイミングの不確実性を分析に組み入れて、改革の遅れがもたらす厚生損失を定量的に示した。
\citet{Kitao2015JEDC} quantitatively explores the fiscal impacts of demographic change in Japan, focusing on the pension reform.
She shows that the consumption tax rate needed to maintain the public finances would rise to a maximum of nearly 50\% without pension reforms, and that raising the pension starting age together with a 20\% benefit reduction would result in a consumption tax rate raise of less than 30\%.
Furthermore, \citet{Kitao2018RED} focuses on the pension reform and incorporated uncertainty about the timing of beginning reforms to reduce pension benefits into the analysis to illustrate quantitatively the welfare losses that would result from a delay in reforms.

%一方、特に労働力の減少の影響に着目した研究も存在する。
%たとえば、■İmrohoroğlu, Kitao, and Yamada (2017) は、人口減少下の日本において、guest workers（外国人労働者）の流入による労働力増加が財政の維持可能性やnative Japaneseの厚生変化へ与える効果を分析した。
%なお、彼らはguest workersを日本で10年間働いて社会保障給付は受け取らずに帰国すると仮定する。
%そのうえで、彼らはguest workersの技能水準の高低に区別したシナリオ分析を行い、特に高技能のguest workers導入は将来の消費税率を引き下げ、将来のnative Japaneseの厚生を改善することを示した。
%詳細なミクロデータに基づいて家計間の異質性や社会保障制度を設定して財政維持可能性を分析した研究もある。
%特に■İmrohoroğlu, Kitao, and Yamada (2019) は、年金に加えて医療・介護保険制度の改革も考慮したうえで、労働生産性と女性の労働力参加の向上が財政の維持可能性の改善により高い効果があることを示した。

Some studies find implications that focus on the relationship between labor market policies and fiscal sustainability.
\citet{IKY2017EI} measure the effects of labor force growth due to the immigration of guest workers on fiscal sustainability and welfare changes among native Japanese people in a shrinking population. 
They suppose that guest workers work in Japan for a decade and return to their own country without receiving social security benefits. 
They conduct a scenario analysis distinguishing between high and low skill levels of guest workers, and show that an influx of high-skilled guest workers lowers the consumption tax rate in particular and improves the welfare of native Japanese people in the future.
Moreover, \citet{IKY2019JEA} find that not only encouraging women's labor force participation, but also increasing women's labor productivity, combined with social security reforms, including pensions and health and long-term care insurance, are effective for improving fiscal sustainability.
In their research, they develop a detailed overlapping generations model based on Japanese micro data, which incorporates heterogeneity in household income and labor supply.

\subsection{Inequality, gender, and employment type}

%近年、社会保障改革の影響が家計の所得・資産の水準や属性によって異なることに着目した研究が蓄積されている。

%■McGrattan and Prescott (2017, 2018)は米国の家計を資産階層のミクロデータに基づいて4タイプのエージェントに区別し、賦課方式（PAYG）に基づく公的年金制度を個人積立方式に移行した場合の厚生への影響を分析した。
%その結果、著者らは積立方式への移行には社会厚生を引き上げる効果があることを示した。
%これらの研究は、累進的な労働所得税率を設定することで、詳細に家計のライフサイクルと公的年金の保険料・給付の関係を捉えている。
%さらに、これらの研究では、通常の民間部門と法人税の課税対象外である政府・個人事業主部門を想定した2部門の企業を設定し、さらに有形資産だけでなく無形資産もモデルに組み込むことで、より現実的な想定に基づく検証を行っている。
%%% ★西山先生の論文も引用する？？：異質エージェント世代重複（脚注に入れるかどうかは要検討）
%%■McGrattan, Miyachi, and Peralta-Alva (2019) は、■McGrattan and Prescott (2017) に基づいて、エージェントを資産（wealth）階級で4タイプに区別したモデルを用いて、日本の社会保障改革 (年金、医療・介護保険) に関する分析を行った。

Recent research suggests that the impact of social security reform varies by household income and wealth levels, gender, and other characteristics.
\citet{MP2017QE, MP2018JA} analyze the impact of a public pension system reform from a PAYG system to a privately funded system using an OLG model with household heterogeneity characterized by four types in terms of their wealth class.
They quantified the degrees to which a change to a privately funded system has improved social welfare for heterogeneous households with income and wealth inequality.
To capture the gains and losses between public pension contributions and benefits over the life cycle with higher accuracy, they incorporated a progressive labor income tax rate into their model. 
They also incorporate into their model two firm sectors: the corporate sector and a household business (self-employment) sector, and two assets: tangible and intangible assets. 
These detailed settings using macro and micro data allow the model to examine various practical policy simulations focused on the inequality between heterogeneous households.
Furthermore, following \citet{MP2017QE}, \citet{MMP2018IMF} apply the model to the Japanese case and quantified studied social security reforms including those of pensions and medical and LTC insurance, by classifying types of agents into four wealth classes.

\begin{figure}[tb]
 \centering

 \includegraphics[width=100mm]{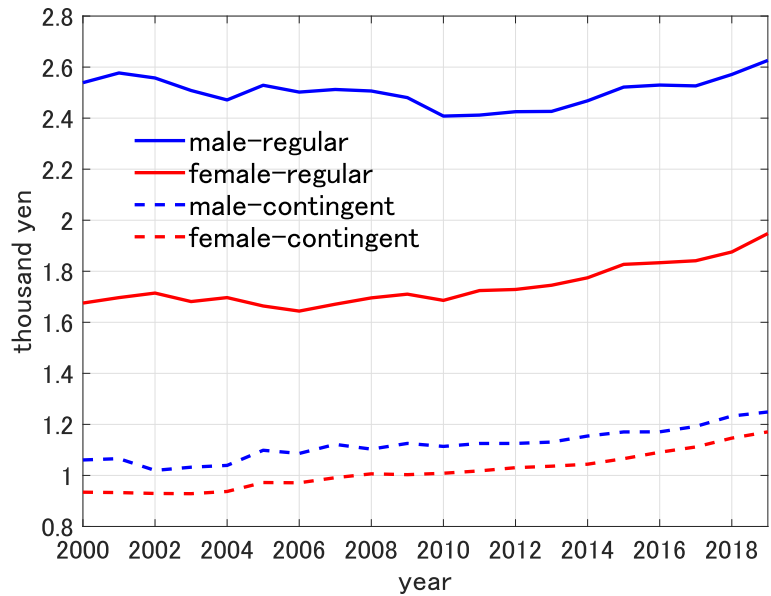}

 \vspace{-0.3\baselineskip}
 \caption{Time series in nominal hourly wages by type}
\label{fig:Intro_wage_change}
\end{figure}

%しかし、日本ではジェンダーや雇用形態が所得や富の不平等を決定づける重要な要因であることが指摘されている（e.g.■Moriguchi 2017）。
%■Moriguchi 2017は、日本では長期雇用と手厚い社会保障制度が適用されるregular workersとそうではないcontingent workersの所得や資産の差が大きく、近年contingent workersの割合が増えていることが格差の拡大に貢献していることを議論している。
%実際、contingent workersの割合は、2000年に男性は約10\%、女性は約45\%であったが、2019年には男性は約20\%、女性は約52\%へと上昇している（総務省統計局、労働力調査）。
%さらに、賃金のジェンダー間のギャップ、雇用形態間のギャップは過去から残存し続けている。Figure \ref{fig:Intro_wage_change}は2000年から2019年の性別・雇用形態別の名目時間当たり賃金の変化を描いている（厚生労働省、賃金構造基本統計調査）。この図が示すように、日本においては労働者のタイプ間の賃金ギャップにはほとんど変化が見られない。そのため、日本で家計の異質性を着目した分析を行うには、性別と雇用形態で区別することが重要だと考えられる。
However, several studies point out that gender and employment types are essential determinants of income and wealth inequality in Japan's labor market \citep[e.g.,][]{Moriguchi2017}. 
\citet{Moriguchi2017} argues that the gaps in income and wealth between regular workers, who are employed long-term (or until their retirement age in most cases) and receive generous social security benefits, and contingent workers, who are not taken those benefits, is significant in Japan. Moreover, she shows the increasing share of contingent workers has contributed to the expansion of this inequality in recent years.
The share of contingent workers has risen from about 10\% for males and 45\% for females in 2000 to about 20\% for males and 52\% for females in 2019.
\footnote{The Statistics Bureau of the Ministry of Internal Affairs and Communications, the Labor Force Survey.}
Furthermore, the wage gaps between gender and employment types have remained for several decades. Figure \ref{fig:Intro_wage_change} depicts the time series in nominal hourly wages by gender and employment types from 2000 to 2019, which is based on the Basic Wage Structure Survey by the Ministry of Health, Labour and Welfare.
\footnote{"Regular workers" are regarded as full-time workers employed directly by an employer. On the other hand, "Contingent workers" include part-time workers, workers on fixed-term contracts, and dispatched workers.
These precise definitions differ depending on the statistics and context.
This paper considers "general workers" (\textit{ippan roudousya}) and "part-time workers" (\textit{tanjikan roudousya}) in the Basic Survey on Wage Structure to correspond to regular workers and contingent workers.}
As this figure shows, there has been little change in relationship of the wage gaps between these four types of workers for long period. 
Accordingly, we classify agents in terms of gender and employment types in order to conduct an analysis focusing on household heterogeneity in Japan.

%%%%%%%%%%%%%%%%%%%%%%%%%%%%%%%%%%%%%%%%%%%%
%%%%%%%%%%%%%%%%%%%%%%%%%%%%%%%%%%%%%%%%%%%%% 

%実際、性別・雇用形態に基づく異質性に着目した、日本の社会保障、財政や税制に関する研究が行われている。
%■注■　もちろん日本以外でも、性別による違いに着目した研究は蓄積されている。たとえば、■Braun, Kopecky, and Koreshkova (2017)は男女で異なる米国の老年期に直面する医療・介護支出、配偶者との死別等のリスクを捉えたうえで、老年所得補償制度の望ましい水準を分析している。
Actually, studies regarding Japan's social security, fiscal and taxation systems have focused their analysis on heterogeneity based on gender and employment type.
%たとえば、■Kitao and Mikoshiba (2020JJIE) では労働参加率、雇用形態、労働生産性、平均余命が異なる男女を含む世代重複モデルを用いて、労働市場の変化が財政の持続可能性に及ぼす影響を検証する。
%その結果、彼女たちは特に女性や高齢者の労働参加率と労働生産性の向上が、財政の改善に重要であることを示した。
%また、■Kitao and Mikoshiba (2022FLFP) は、日本が、主に結婚家計の女性に対して適用する優遇的な税制（自身の所得に応じた税額控除等）の影響に着目する。そして彼女らは、女性の労働力参加を引き下げているのではないか、労働所得の低いcontingent workerとしての就業を促進しているのではないか、という問いについて議論する。
%その問いを明らかにするために、彼女たちは性別、雇用形態、教育達成（スキル）の異なる人々を想定した構造ライフサイクルモデルを構築し、それらの税制が無かった場合のカウンターファクチュアル・シミュレーションを行った。
%その結果、それらの税制が無かった場合には女性の労働力参加は12.5\%、所得は22.7\%上昇しうることを示した。
%特に所得の上昇が顕著な要因として、女性の労働力参加と雇用形態の選択を妨げる税制がなくなることでregular workerとして就業する者が増加することを指摘する。
%さらに、これのよって政府の税・社会保険料収入も増加することも示したうえで、政府はこの税制改革により、現役世代の厚生を引き下げることなく税収を増加できる可能性について議論している。
For example, \citet{KSMM2020JJIE} developed an OLG model that incorporates males and females with different employment type, labor participation rates, labor productivity, and life expectancy to examine the impact of labor market changes on fiscal sustainability.
They showed that an increase in labor participation rates and labor productivity, especially of females and the elderly, could be effective for contributing to the transition away from budget deficits.
\citet{KSMM2022FLFP} quantified the effects of tax incentive policies mainly applied to married women (i.e., tax credits for their spouse's income). They examined the extent to which the tax incentive treatment reduces women's labor force participation and compromises them to work as low-income contingent workers.
Their counterfactual simulations show that females' labor force participation and income could increase by 12.5 and 22.7\%, respectively, without these tax incentives. 
They suggest that removing taxation distortions to their labor force participation and choice of employment types would increase the number of females working as regular workers, which is expected to bring a marked rise in income for their households. 
They also discussed that the government might increase tax revenues through this tax reform without reducing the welfare of the current generations.

Of course, many researchers have also focused on gender and employment type differences in various countries outside Japan.
For example, \citet{Braun&KK2017} discuss the effective of old-age public income compensation programs by analyzing the risks of medical and LTC expenditures, spousal bereavement, and other risks faced in old age in the US that differ for males and females. Furthermore, \citet{mukoyamaCyclicalPartTimeEmployment2021} highlight the importance of distinguishing between full-time workers and part-timers in the labor market.
They argue that the share of part-timers increased after the Great Recession in the U.S. and that the most important contributor was the transition from full-timers to part-timers.
They also suggest segmenting the labor market for full-timers and part-timers is essential to reproduce the labor market dynamics during the Great Recession.

%そこで、われわれは日本における家計の所得・資産格差の重要な決定要因の1つである性別と雇用形態に着目し、われわれは現在世代と将来世代の間の社会保障改革のインパクトの異質性に加えて、世代内における家計のタイプごとの影響の異質性を定量的に明らかにする。加えて、社会保障改革のコストとベネフィットについて議論し、政策的な含意を導く。

Accordingly, we classify households from gender and employment types, which might be essential determinants of Japan's household income and wealth inequality, as well as from generations.
We quantitatively analyze the heterogeneity of the impact of social security reforms between current and future generations, as well as the heterogeneity of the effects by household type distinguished by gender and employment types within a generation.
In addition, the paper discusses the costs and benefits of social security reform based on the results of the quantitative analysis and derives policy implications.

\subsection{Savings and medical and LTC expenditures in retirement}
%上記とは別の視点から医療・介護に着目した研究も数多く行われている。
%特に、引退期の貯蓄が通常のライフサイクルモデルによる予測よりも一般的高いこと（retirement saving puzzle）の要因として、医療・介護への支出に着目した研究がある。
%たとえば、■DeNaldi, French and Jones (2010) は、高齢期に生じうる医療支出などを賄うために予備的貯蓄という側面から高齢者世帯の貯蓄行動を説明する。
%加えて、■Kopecky and Kreshkova (2014) は、医療支出に加えて介護支出が高齢者の予備的貯蓄の高さを説明する要因として重要であることを指摘する。
%また、米国の低所得者向け医療保険であるMedicaidを拡大することで、将来世代の厚生が改善される可能性も示した。
%さらに、■Braun, Kopecky and Koreshkova (2017) は医療・介護支出に加えて、配偶者の死や長寿などといったより広い高齢期のリスクを考慮して、米国のミーンズテスト付き社会保障の拡大が厚生の改善に結びつく可能性を示した。
Turning to medical and LTC costs from different perspectives, 
studies have been conducted on the retirement savings puzzle, which indicates that the actual medical and LTC expenditures are higher than those derived using the standard life cycle model. 
For example, \citet{DeNardietal2010} explain the saving behavior of elderly households in the context of precautionary savings to finance the  medical expenditures that may arise during old age.
In addition, \citet{KKLTC2014} find that the LTC expenditures account for a large portion of precautionary savings among the elderly.
They also show that expanding Medicaid, which is the U.S. health insurance program for low-income households, could improve the welfare of future generations.
Moreover, \citet{Braun&KK2017} show that expanding social security with a means test could improve welfare by taking into account aging risks, such as death or the longevity of a spouse.

\begin{figure}[t]%[htp]
 \centering
 \includegraphics[width=105mm]{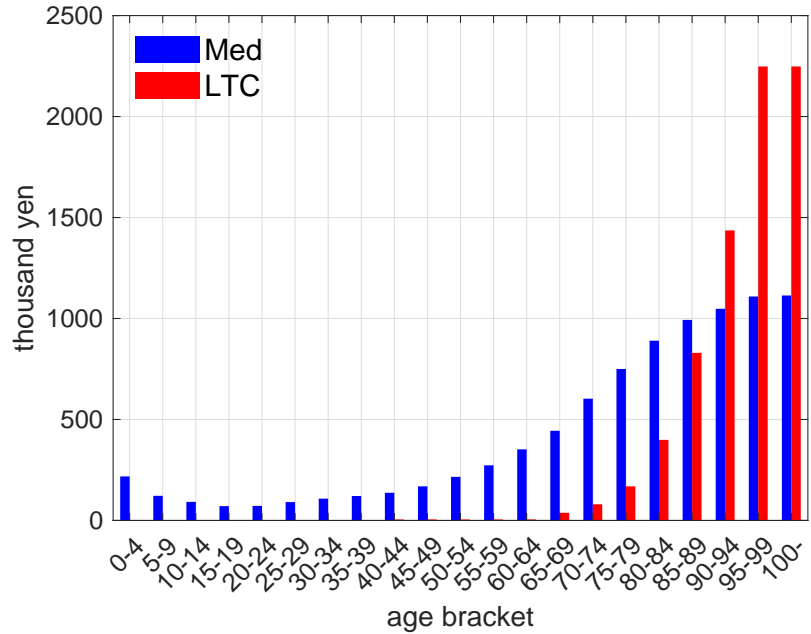}

 \vspace{-0.3\baselineskip}
 \caption{Per capita medical and LTC expenditure}
 \label{fig:Med-LTC_expend}
\end{figure}

%日本における、平均的な医療・介護支出は、図\ref{fig:Med-LTC_expend}に示されている。
%この図は、■Iwamoto and Fukui (2018) で示された年齢階級別の1人当たり医療費と介護費の金額に基づいている。
%\footnote{■Iwamoto and Fukui (2018)は厚生労働省の資料（「医療保険に関する基礎資料」「介護給付費等実態調査」）に基づいて算出している。なお、この数字は自己負担分と保険負担分の合計金額である。}
%この図は、高齢期において医療・介護費用が多くなることを示している。
%特に、介護費用は80代以降に急激に上昇する。
%老年期に予想される多額の支出は、若年期の貯蓄・消費行動に大きな影響を与える。
%加えて、医療・介護の自己負担率の引き上げは、その影響をさらに大きくすると我々は考える。
For Japan, \citet{IwamotoFukui2018} show average medical and LTC expenditures.
\footnote{
\citet{IwamotoFukui2018} is calculated on data from the Ministry of Health, Labour and Welfare (Basic Data on Medical Insurance and Statistics of Long-term Care Benefit Expenditures). This is the sum of the copayment and insurance contribution.}
Figure \ref{fig:Med-LTC_expend} depicts the per capita medical and LTC expenditures by age group. As the figure shows, medical and LTC costs are higher for the elderly.
In particular, LTC costs raise rapidly after the age of 80.
Increasing the expected spending in old age must influence the amount of savings and consumption  at a young age.
Similar to this, increases in the copayment rates for medical and LTC expenditures could have a much greater impact on them at a young age. 
We take the impact of such cost-up into account in our study, by incorporating both the medical and LTC costs into our model.

%%%%%%%%%%%%%%%%%%%%%%%%%%%%%%%%%%%%%%%%%%%%
%%%%%%%%%%%%%%%%%%%%%%%%%%%%%%%%%%%%%%%%%%%%

%上記ふまえ、本論文では社会保障改革が家計の消費・貯蓄行動や厚生に及ぼす影響の異質性に着目した政策シミュレーションを行うために、性別・雇用形態で区別した4タイプのエージェントを想定した世代重複モデルを構築する。
%Based on the above, our paper develops an OLG model with four types of agents distinguished by gender and employment type to conduct policy simulations focusing on the heterogeneity of the impacts of social security reform on household consumption and saving behavior and welfare.

%%%%%%%%%%%%%%%%%%%%%%%%%%%%%%%%%%%%%%%%%%%%%%%%%%%%%%%%%%%%%%%%%%%%%%%%%%%%%%%%%%%%
\section{Model}
\label{sec:MODEL}

%本節では、我々が定量分析で用いるモデルを記述する。
%まずは社会保障制度を描写したモデルを示し、次に家計、企業のモデルを説明する。
%さらに、社会保障制度を運営する政府の予算制約等について述べたうえで、最後に均衡条件を定義する。
In this section, we describe the model used in our quantitative analysis. 
First, we present the settings that describe the social security system. 
Next, we set up households and firms, and describe the budget constraints of the government managing the social security system. Finally, we define the equilibrium conditions.

%前のsectionで述べた通り、我々のモデルは■McGrattan and Prescott (2017) に基づいて、日本の特徴を捉えるために拡張したdeterministic OLG modelである。
%家計部門は性別と雇用形態で区別された4タイプのエージェントで構成される。
%企業部門は、法人企業と個人事業主の2部門で構成され、有形資産と無形資産が考慮される。
%政府は、社会保障（年金、医療、介護）財源調達と給付を担う。
As mentioned in the previous section, our model is a deterministic OLG model following \citet{MP2017QE} modified to fit Japan's original system.
The households consist of four types of agents distinguished by gender and employment type.
The firms are divided into two sectors, 
specifically the corporate sector, and the household business (self-employment) sector,
the capital of which is formed from tangible and intangible assets.
The government runs the social security systems (pensions, medical insurance, and LTC insurance) as well as conducting finance by issuing government bonds and provides benefits for households.

%われわれのモデルは、perfect foresightで行動する家計と、完全競争市場において規模に関して収穫一定の生産技術を持った各部門の代表的企業で構成されている。
%また政府はcredible fiscal policyを実施する。
%時間は離散的であり、モデル上の1期間（$t$）は1年である。
%加えて、我々モデルでは経済は長期において一定のlabor-augmentedなproductivityの成長率$γ$と、一定の人口成長率$n$を持つbalanced growth path上にあると仮定する。
We suppose that households make decisions with perfect foresight and that representative firms in both sectors produce with a constant return to scale technology under perfect competition.
The government implements a credible fiscal policy.
Time is discrete, and one period in the model is one year.
We also suppose that the economy is on a balanced growth path with a constant labor-augmented productivity growth rate $\gamma$ and a constant population growth rate $n$ in the long-run.

\subsection{Social security system}

\subsubsection{Demography}
First, let $t$ and $j$ denote the time period and the age of the agents at the maximum age $J$, respectively.
$g \in \left\{ m,\ f \right\}$ represents gender. $m$ and $f$
denote male and female. 
$s_{t}^{j,g}$ represents the probability that an agent of gender $g$ entering the economy at age $j = 1$ will survive until the next period.
The unconditional survival probability at age $j$ in period $t$ is given by
\begin{align}
S_{t}^{j,g} = \prod_{k = 1}^{j}s_{t + (k - j)}^{k,g}.
\end{align}
We also denote as $n_{t}^{j,g,h}$ the size of the population at each group, where $h \in \{regular, contingent\}$ signifies the employment types and \textit{regular} and  \textit{contingent} represent a regular worker and a contingent worker, respectively.
The size of the new cohort $n_{t}^{j,g,h}$ for $j=1$ grows at rate
$\gamma_{t,n}$.

\subsubsection{Pension system}
Following \citet{KSMM2020JJIE}, we set up the pension benefits that households are supposed to receive  after they reach normal retirement age $J^{R}$.
$p_{t}^{j,g,h}$ is the public pension benefit received at age $j$ in
period $t$ where $g$ denotes gender and $h$, employment type. The public pension benefit is given as
\begin{align}
p_{t}^{j,g,h} = \kappa_{t}\frac{W_{t}^{j,g,h}}{J^{R} - 1},
\end{align}
where $\kappa_{t}$ is the replacement rate of the past average
earnings.
The accumulated labor income earned by  heterogeneous households during their lifetime,
$W_{t}^{j,g,h}$, is defined recursively as
\begin{align}
W_{t}^{i,g,h} = \left\{ \begin{matrix}
w_{t}l_{t}^{j,g,h}e^{j,g,h} + W_{t - 1}^{j - 1,g,h}, & \text{if}\,\, j < J^{R}, \\
W_{t - 1}^{j - 1,g,h}, & \text{otherwise}, \\
\end{matrix} \right.
\end{align}
where $w_{t}$ is the wage rates in period $t$. $l_{t}^{j,g,h}$ and $e^{j,g,h}$ are the labor supply, or working hours, and labor productivity of the households differentiated by age, gender, and employment type, respectively.
Therefore, the aggregate public pension expenditure provided by the government in each year can be defined as
\begin{align}
P_{t}^{g} = \sum_{j = J^{R}}^{J}n_{t}^{j,g,h}p_{t}^{j,g,h}.
\end{align}

\subsubsection{Medical and LTC expenditures}
In line with \citet{Kitao2015JEDC, Kitao2018RED}, we introduce the total healthcare cost, which consists of both medical costs $m_{j,t}^{H}$ and LTC costs $m_{j,t}^{L}$, faced by every household each period.
The fractions $\lambda_{j,t}^{H}$ and $\lambda_{j,t}^{L}$ represent the copayment rates for medical and LTC expenditures.
The remaining expenditures are covered by public medical insurance and LTC insurance.
The aggregate healthcare cost is the sum of households' copayments and government contributions $M_{t}^{g}$ and is defined as
\begin{align}
M_{t}^{g} = \sum_{j = 1}^{J}\left\lbrack \left( 1 - \lambda_{j,t}^{H} \right) m_{j,t}^{H} + \left( 1 - \lambda_{j,t}^{L} \right) m_{j,t}^{L} \right\rbrack n_{t}^{j,g,h}.
\end{align}

\subsection{Households}
\subsubsection{Interest rate of assets and government bonds}

Following \citet{BJ2015JEDC}, \citet{Kitao2015JEDC, Kitao2018RED}, and \citet{MMP2018IMF}, 
we suppose that households allocate a fraction $\psi_{t}$ of their assets to government debt and the rest to corporate equity.
Thus, the after-tax gross return on personal assets is given by
\begin{align}
1 + r_{t} = \psi_{t}\left( 1 + i_{t}^{d} \right) + \left( 1 - \psi_{t} \right)\left( 1 + i_{t}^{k} \right),
\end{align}
where $i^{d}$  and $i^{k}$ are the interest rate on government bonds, and the interest rate on private capital, respectively.
In our setting, the former is exogenously, while the latter is endogenously determined.

\subsubsection{Optimization problem}
For households, a single period of risk-free assets may be traded.
This asset is a composite of investment in corporate capital and a holding of government debt, paying gross interest $r_{t}$ after taxes.
We also suppose that borrowing against future income or transfers is not allowed and that assets must be non-negative.
The state vector of an agent $x\,( = \{ j,a_{t},g,h\})$ consists of age $j$, assets $a_{t}$, gender $g$, and employment type $h$.
The agent chooses the optimal path of consumption $c_{t}$, assets $a_{t + 1}$, and labor supply $l_{t}$ to maximize the lifetime utility, where $l_{t}$ indicates the working hours, or intensive margin, and $l_{t} \geq 0$. The labor supply of households after reached their retirement age $J_{R}$ is assigned as $l_{t} = 0$.
The problem is solved recursively, and the value function $v_{t}(x)$ without policy uncertainty is defined as
\begin{align}
& v_{t}\left( j,a_{t},g,h \right) = \max_{a_{t + 1},c_{t},l_{t} > 0} \left\{u\left( c_{t},l_{t} \right) + \beta s_{t}^{j,g}v_{t + 1}\left( j + 1,\ a_{t + 1},g,h \right)\right\},
\\
& \text{s.t.}\text{\ \ }\left( 1 + \tau_{ct} \right)c_{t + 1} + \lambda_{j,t}^{H} m_{j,t}^{H} + \lambda_{j,t}^{L}m_{j,t}^{L} + s_{t + 1}^{j + 1,g}a_{t + 1} \notag 
\\
& \hspace{1.5em} = \left( 1 + r_{t} \right)a_{t} + y_{t} - T_{t}^{w}\left( y_{t} \right) + T_{t,g,h}^{p} + D_{1t} + D_{2t},
\label{eq:constraint_1}
\\
& \hspace{0.6em} y_{t} = w_{t}l_{t}e^{j,g,h},
\label{eq:constraint_2}
\end{align}
where the value function $v_{J + 1}$ at the final age is equal to zero.
Agents can allocate a unit of disposable time to the labor supply and leisure in each period.

We suppose that agents know with certainty the profile of the labor efficiency $e^{j,g,h}$ that depends on the age, gender, and employment type.
Private capital and government bonds are shares of ownership in an asset that pay out at the time of their retirement, and the return on assets is provided to the still living members of their same cohort. 
Thus, the return depends not only on the size of the capital and bond but also on the probability of survival.

For the left-hand side of equation \eqref{eq:constraint_1}, we show that agents only pay out of pocket for medical and LTC expenditures and receive healthcare services; following \citet{BJ2015JEDC}, 
%Braun and Joines (2015) 
this is not reflected in the utility function as in normal consumption.
That is, we suppose that health care costs are always incurred by households, although the amount varies by age and gender. 
The budget constraint of households is also set to be consistent with the fact that health care costs are not subject to consumption tax.
As for the right-hand side of equation \eqref{eq:constraint_1}, the first term is the after-tax gross income on the assets $\left( 1 + r_{t} \right)a_{t}$ explained above, and the next term $y_{t}$ is the labor income, which indicates that households receive an income transfer of pension benefit $T_{t}^{p}$ while being subject to the labor income tax and social security contribution $T_{t}^{w}$.
In addition, $D_{1t}$ and $D_{2t}$ are the dividends received from firms in Sectors 1 and 2, respectively.
Here, Sector 1 denotes the corporate sector, which is subject to corporate income tax, while Sector 2 represents the household business sector, which is not subject to corporate income tax.
Equation \eqref{eq:constraint_2} denotes the labor income in period $t$. 
The labor income $y_{t}$ is given by the market wage rate $w_{t}$, endogenously chosen working hours $l_{t}$, and labor productivity determined by age, gender, and employment type.

\subsection{Firms}
Following Mc Grattan and Prescott (2017) and Mc Grattan et al (2019), we set up two sectors of production in our economy.
Sectors 1 and 2 produce intermediate goods $Y_{1t}$ and $Y_{2t}$, respectively. 
Sector 1 refers to the corporate sector, while Sector 2 is the household business sector.
The aggregate production function of the composite final good is given by
\begin{align}
Y_{t}\, = \, Y_{1t}^{\eta_{1}}\, Y_{2t}^{\eta_{2}},
\label{eq:production_func}
\end{align}
where $\eta_{1} = 1 - \eta_{2}$. The setting indicates constant returns to scale, and both parameters, $\eta_{1}$ and $\eta_{2}$, are positive. 
In addition, the production function in Sector $i$ ($i = 1, 2 $) is formed by the Cobb-Douglas function with inputs of tangibles denoted as $T$ and  intangibles denoted as $I$ capital stocks, that is, $K_{iTt}$  $K_{iIt}$, and labor $L_{it}$, as below.
\begin{align}
Y_{it} = A_{t}K_{iTt}^{\theta_{it}}K_{iIt}^{\theta_{iI}}(\Omega_{t}L_{it})^{1 - \theta_{it} - \theta_{iI}},\quad i = 1, 2.
\end{align}
The levels of TFP and labor-augmented technology growth rate in period $t$ for Sector $i$  are, respectively, $A_{t}$ and $\Omega_{t}$, which grow at rates of $\gamma_{A}$ and $\gamma_{\Omega}$.
\begin{align}
A_{t + 1} & = (1 + \gamma_{A})A_{t}, \\
\Omega_{t + 1} & = (1 + \gamma_{\Omega})\Omega_{t}.
\end{align}
The tangible and intangible capital stocks depreciating at certain rates, $\delta_{iT}$, $\delta_{iI}$, in Sector $i$ are given as
\begin{align}
K_{iT,t + 1} & = (1 - \delta_{iT})K_{iTt} + X_{iTt}, \\
K_{iI,t + 1} & = (1 - \delta_{iI})K_{iIt} + X_{iIt},
\end{align}
where $X_{iTt}$ and $X_{iIt}$ are tangible and intangible investments for $i = 1,\ 2$, respectively.
In addition, from the resource constraint, the output $Y_{t}$ and gross domestic product ($\text{GD}P_{t}$) are expressed as
\begin{align}
Y_{t} & = C_{t} + X_{Tt} + X_{It} + G_{t}, \\
\text{GD}P_{t} & = C_{t} + X_{Tt} + G_{t},
\end{align}
where
$X_{Tt} = \sum_{i}X_{iTt},\, X_{It} = \sum_{i}X_{iIt}$.

\subsection{Government}

The government in our economy collects revenue through consumption taxes, labor income taxes, social security premiums, corporate and dividend taxes, and the issuance of risk-less debt $B_{t + 1}$.
Its revenue is used to purchase goods and services $G_{t}$, to pay the principal and interest on debt $(1 + i^{d})B_{t}$, and for public pension benefits $P_{t}^{g}$ and medical and LTC insurance benefits $M_{t}^{g}$.

\paragraph{Labor income tax and social security premium}
We suppose a labor income tax based on a progressive taxation system, which includes a social security premium. 
The details of the progressive taxation system will be explained in the next section.

\paragraph{Corporate income taxes and dividend taxes}
The accounting profit of Sector 1, the corporate sector, in period $t$ is given by
\begin{align}
\Pi_{1t} = p_{1t}Y_{1t} - w_{t}L_{1t} - X_{1It} - \delta_{1T}K_{1Tt},
\end{align}
where $p_{1t}$ is the relative prices of intermediate goods to final
goods in Sector 1.
\footnote{
We suppose the final goods as the numeraire, that is their price is the unity.}

%%\footnote{★有形資産では\deltaを考慮して減価償却するけど、無形資産は償却しないことを説明する必要あるか。詳しくはMcGrattanを見よ\}→書かなくてOK

Let us now write the corporate tax of Sector 1 as $\tau_{1t}^{\pi}$; the dividend of the shareholders in this sector $D_{1t}$, can be
expressed by the following equation:
\begin{align}
D_{1t} = \left( 1 - \tau_{1t}^{\pi} \right)\Pi_{1t} - K_{1T, t + 1} + K_{1Tt}.
\end{align}
Similarly, the dividend $D_{2t}$, for Sector 2, the household business sector, is given by
\begin{align}
D_{2t} = \Pi_{2t} = p_{2t}Y_{2t} - w_{t}L_{2t} - X_{2It} - \delta_{2T}K_{2Tt},
\end{align}
where $p_{2t}$ is the relative prices of intermediate goods to final
goods in Sector 2.

In our economy, the profits of the corporations in Sector 1 are taxed at
the rate $\tau_{1t}^{\pi}$. And the dividends $D_{It}$ of Sector $i$ are taxed at the rate $\tau_{It}^{d}$.

\paragraph{Government debt}
From the above public expenditures and revenues, the equation for government debt can be derived as follows:
\begin{align}
B_{t + 1} = & B_{t} + i^{d}B_{t} + G_{t} + P_{t}^{g} + M_{t}^{g} \notag 
\\ 
& - \sum_{j,g,h}^{}n_{t}^{j,g,h}T_{t}^{w}\left( w_{t}l_{t}^{j,g,h}e^{j,g,h} \right) \tau_{t}^{c}C_{t} - \tau_{1t}^{\pi}\Pi_{1t} - \tau_{1t}^{d}D_{1t} - \tau_{2t}^{d}D_{2t}.
\end{align}

\subsection{Equilibrium conditions}
The labor income progressive tax and capital tax rates and the ratio of government consumption to GDP are given exogenously, while the government debt, private capital interest and wage rates, and consumption tax rate are determined endogenously so that the market equilibrium is complete.

A competitive equilibrium, for a given sequence of demographics, TFP levels $\{ A_{t}\}_{t = 1}^{\infty}$, 
and fiscal variables
$
\left\{
G_{t}, B_{t}, \tau_{1t}^{\pi}, \tau_{1t}^{d}, \tau_{2t}^{d}, \{\tau_{wt}\}_{j,g,h}, \{ p_{j,g,h,t}\}_{j,g,h}, 
\{ m_{j,g,h,t}^{H} \}_{j,g,h}, \{ m_{j,g,h,t}^{L} \}_{j,g,h}
\right\}
$,
is a sequence of
\begin{itemize}
\item
  Households' choices
  $\left\{\{ c_{j,g,h,t},l_{j,g,h,t},a_{j,g,h,t}\}_{j,g,h}\right\}_{t = 1}^{\infty}$,
\item
  Consumption tax rates
  $\{\tau_{ct}\}_{t = 1}^{\infty}$,
\item
  Wage rates $\{ w_{t}\}_{t = 1}^{\infty}$, 
  interest rates $\{ r_{t}\}_{t = 1}^{\infty}$, and
\item
  Aggregate variables
  $\{K_{1Tt},K_{2Tt},K_{1It},K_{2It},L_{1t},L_{2t}\}_{t = 1}^{\infty}$.
\end{itemize}

%%%%%%%%%%%%%%%%%%%%%%%%%%%%%%%%%%%%%%%%%%%%%%%%%%%%%%%%%%%%%%%%%%%%%%%%%%%%%%%%%%%%
\section{Calibration}\label{sec:CALIB}
In this section, we describe the procedure for calibrating and parameterizing the model and we explain the sources of data used for the calibration. The quantitative results of the transition path for our policy simulations are discribed in the next section.

We conduct the calibration of the steady state and transition path in our model based on the three steps below, following \citet{MP2017QE,MP2018JA}.
The first step is to calculate the initial steady state. 
To be more precise, we calculate the initial steady state by setting parameters based on Japan's demographics and economic data, such as the National Accounts (JSNA). That is, we use the demographic structure as of 2015 and the average value of the JSNA from 2015 to 2019 as the initial steady state data.
The second step is to approximate a balanced growth path. 

Based on the initial steady state calculated in the first step, we suppose balanced growth given the expected growth rate of technological productivity (the TFP growth rate $\gamma_A$ and the labor-augmented technology growth rate $\gamma_\Omega$) and the population growth rate ($\gamma_n$) and derive a balanced growth path of the economy up to 240 years from the base year set as the initial steady state.
The third step is calculating the transition path. 
Based on the balanced growth paths derived in the second step, we calibrate the transition path based on annual projections of population dynamics and other factors. 
In this step, we further conduct policy simulations for retirement age extension and social security reform by changing the condition settings.

In these steps, we parameterize the following two aspects in terms of both macro and micro sides.
First, to fit the actual aggregate data representing the national accounts and fixed asset tables shown in Tables \ref{tab:SNA} and \ref{tab:assettable}, we select the parameters about demographics, household preferences, firm technology, government spending and debt shares, and capital income tax rates. 
Second, for the micro side, we set the levels of labor productivity, labor supply, and progressive labor income tax to match the corresponding data for the population share, average labor income, and average working hours classified by gender and employment type.

\subsection{National income and product accounts}
Table \ref{tab:SNA} shows the national income and product accounts for Japan after some standard adjustments to make the model measures and concepts consistent with the JSNA. 
%脚注追加220913：後述するように、我々はJSNAの各要素の構成比、asset tableのtangebleとintangible assetの構成比と整合的になるように、パラメターをセットしている。表1の2列目（"model"）は、パラメターをセットしてinitial steady stateにおける構成比の計算結果を示している。
\footnote{As described below, we set the parameters consistent with the composition ratios for each category in the JSNA and tangible and intangible assets in the asset table. The second column of Table \ref{tab:SNA} ( "model" ) shows the results of calculating the composition ratios in the initial steady state with the parameters set.}
Table \ref{tab:assettable} reports the capital stock to GDP ratio.
Both tables show averages for the latest five years from 2015 to 2019.

%\input{Tab/table1_pre} %calibrated value なし
%%% 2015～18年の平均値にする。

%%%%%%%%%%%%%%%%%%%%%%%%%%%%%%
\begin{table}[hbtp]
\begin{footnotesize}
\caption{Adjusted JSNA (average from 2015 to 2019) and calibrated value}
\label{tab:SNA}

\begin{center}
\begin{tabular}{lcc}
\hline 
& Data & Model \tabularnewline
\hline 
\textbf{Total adjusted income } & \textbf{1.000} & \textbf{1.000} \tabularnewline
\hline 
\textbf{Labor income } & \textbf{0.509} & \textbf{0.509} \tabularnewline
~~Compensation of employees  & 0.493 & \tabularnewline
~~~~Wages and salaries  & 0.418  & \tabularnewline
~~~~Employers social contributions  & 0.075  & \tabularnewline
~~Households business (70 \% labor income)  & 0.016 & \tabularnewline
% & \tabularnewline
\hline 
\textbf{Capital income } & \textbf{0.491} & \textbf{0.491} \tabularnewline
~~Corporate profits  & 0.160 &  \tabularnewline
~~Households business (30 \% labor income)  & 0.007 &  \tabularnewline
%~~Households operating surplus  & 0.051 \tabularnewline
~~Taxes on production and imports  & 0.082 &  \tabularnewline
~~Less: consumption tax  & 0.039 &  \tabularnewline
~~Less: subsidies  & 0.006 &  \tabularnewline
~~Consumption of fixed capital  & 0.235 &  \tabularnewline
~~Statistical discrepancy  & 0.002 &  \tabularnewline
% & \tabularnewline
\hline 
\textbf{Total adjusted product } & \textbf{1.000} & \textbf{1.000} \tabularnewline 
\hline 
\textbf{Consumption } & \textbf{0.731} & \textbf{0.751} \tabularnewline
~~Private consumption  & 0.539 &  \tabularnewline
~~Less: consumption tax  & 0.039 &  \tabularnewline
~~Government consumption  & 0.192 &  \tabularnewline
\hline 
\textbf{Investment } & \textbf{0.248} & \textbf{0.249} \tabularnewline
~~Gross private investment  & 0.198 &  \tabularnewline
~~~~ corporations  & 0.160 &  \tabularnewline
~~~~ households and NPO  & 0.038 &  \tabularnewline
~~Gross government investment  & 0.050 &  \tabularnewline
~~Changes in inventories  & 0.002 &  \tabularnewline
\hline 
\textbf{Net exports } & \textbf{0.002} & \textbf{0.000} \tabularnewline
\hline 
\end{tabular}
\end{center}

\begin{enumerate}
\item[Notes:]
%Table \ref{tab:SNA} displays the Japanese national income and product accounts, after some of the standard adjustments that make model measurements and concepts consistent with the National Accounts of Japan (JSNA). 
The adjusted GDP is equal to the JSNA GDP after subtracting consumption tax as it is assumed to be levied on private consumption. 
Following \citet{MMP2018IMF}, we categorize income as "labor" or "capital". 
Labor income comprises 53\% of the total adjusted income, and mainly consists of compensation of employees.
Also classified into labor income is 70\% of households' mixed income (including private unincorporated enterprises). 
Meanwhile, capital income includes all other categories of income, including the remaining 30\% of households' mixed income as well as households' net operating surplus (imputed service of owner-occupied dwellings).
The second column ( "model" ) shows the results of calculating the composition ratios in the initial steady state with the parameters set.
\end{enumerate}
\end{footnotesize}
\end{table}
%%%%%%%%%%%%%%%%%%%%%%%%%%%%%%%%%%%%%%%%%%%%%%%%%%% %calibrated value あり

With the 2016 revision of the JSNA, which adopted the 2008 SNA, the capital stock now includes both tangible and intangible capital. 
However, the intangible assets incorporated in our model are a broader concept than those included in the JSNA. 
Therefore, we use the 2018 JIP (The Japan Industrial Productivity) database.
\footnote{
For more details on the JIP database, see the website of the Research Institute of Economy, Trade and Industry (RIETI) (   \url{https://www.rieti.go.jp/en/database/JIP2018/index.html}\quad accessed March 19, 2022).}
The share of private capital in the adjusted GDP is 232\%, of which 68\% is held by private firms and 32\% is held by households and non-profit organizations. 
The capital stock held by the government is 112\% of the adjusted GDP. 
When private capital and public capital are combined, they are equivalent to 343\% of the adjusted GDP. 
\footnote{
We also include land in the capital stock because it is in large part a produced asset associated with real estate development. With land included, the total capital stock is 566\% of adjusted.}
Therefore, we set the capital/output ratio at around three to four in the calibration. 

%%% 2015～18年の平均値にする。TangebleとIntangibleだけ実測値を使って、あとはIMF論文と同じ比率で配分する。

%%%%%%%%%%%%%%%%%%%%%%%%%%%%%%%%%%%%%%%%%%%%%%%%%%%%%
\begin{table}[t]%[hbtp]
\begin{footnotesize}
\caption{Adjusted fixed asset tables with stocks at the end of the period (average from 2015 to 2019)}
\label{tab:assettable}

\begin{center}
\begin{tabular}[t]{lc}
\hline 
\textbf{Tangible capital } & \textbf{5.599 }\tabularnewline
~~Fixed assets  & 3.470 \tabularnewline
~~~~Corporations  & 1.577 \tabularnewline
~~~~Households and NPO  & 0.741 \tabularnewline
~~~~Government  & 1.116 \tabularnewline
~~Land  & 2.128 \tabularnewline
\hline 
\textbf{Intangible capital } & \textbf{0.423 }\tabularnewline
~~R\&D  & 0.199 \tabularnewline
~~Software  & 0.058 \tabularnewline
~~License  & 0.048 \tabularnewline
\hline 
\textbf{Total } & \textbf{6.022 }\tabularnewline
\hline 
\end{tabular}
\end{center}

%\begin{enumerate}
%\item [Notes:] Following \citet{MMP2018IMF}, we take fixed assets including both tangible and intangible assets by the revision of JSNA in 2016 by adopting the 2008 SNA into account as capital stock.
%Private fixed assets amount to 229 \% of adjusted GDP, of which private corporates account for 68 \% and households account for the remaining 32 \%.
%Fixed assets owned by the government amount to 112 \% of adjusted GDP. Together, private and public fixed assets are equal to 340 \% of adjusted GDP.
%We also include land in the capital stock because it is in large part a produced asset associated with real estate development. 
%With land included, the total capital stock is 566\% of adjusted.
%\end{enumerate}
\end{footnotesize}
\end{table}
%%%%%%%%%%%%%%%%%%%%%%%%%%%%%%%%%%%%%%%%%%%%%%%%%%%%5

\subsection{Demographics}
\begin{figure}[tp]
 \centering
\noindent\begin{minipage}[t]{1\columnwidth}%
\begin{center}

\hspace{2em}
\includegraphics[width=66mm, height=63mm]{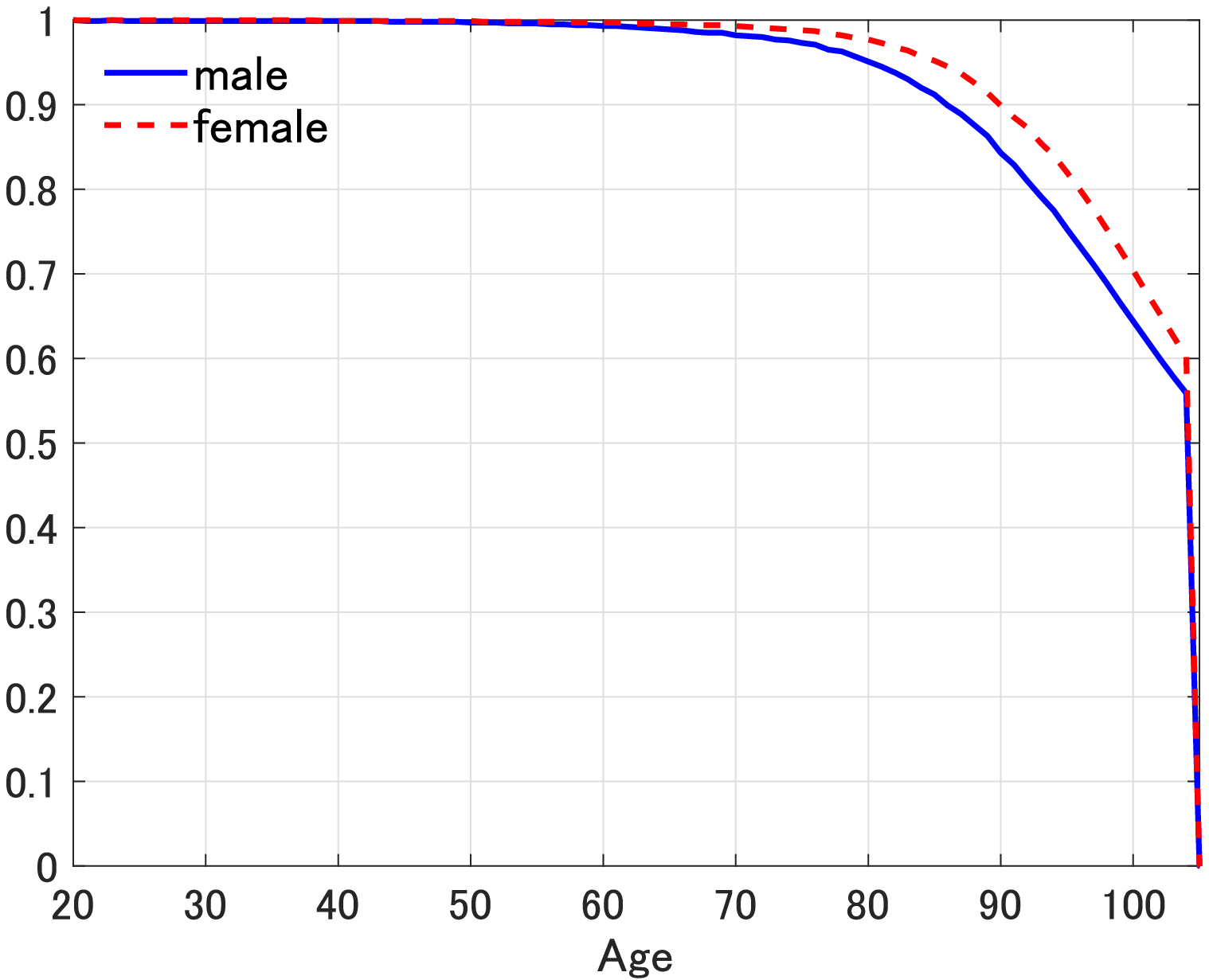}~~~~~~~
\includegraphics[width=66mm, height=63mm]{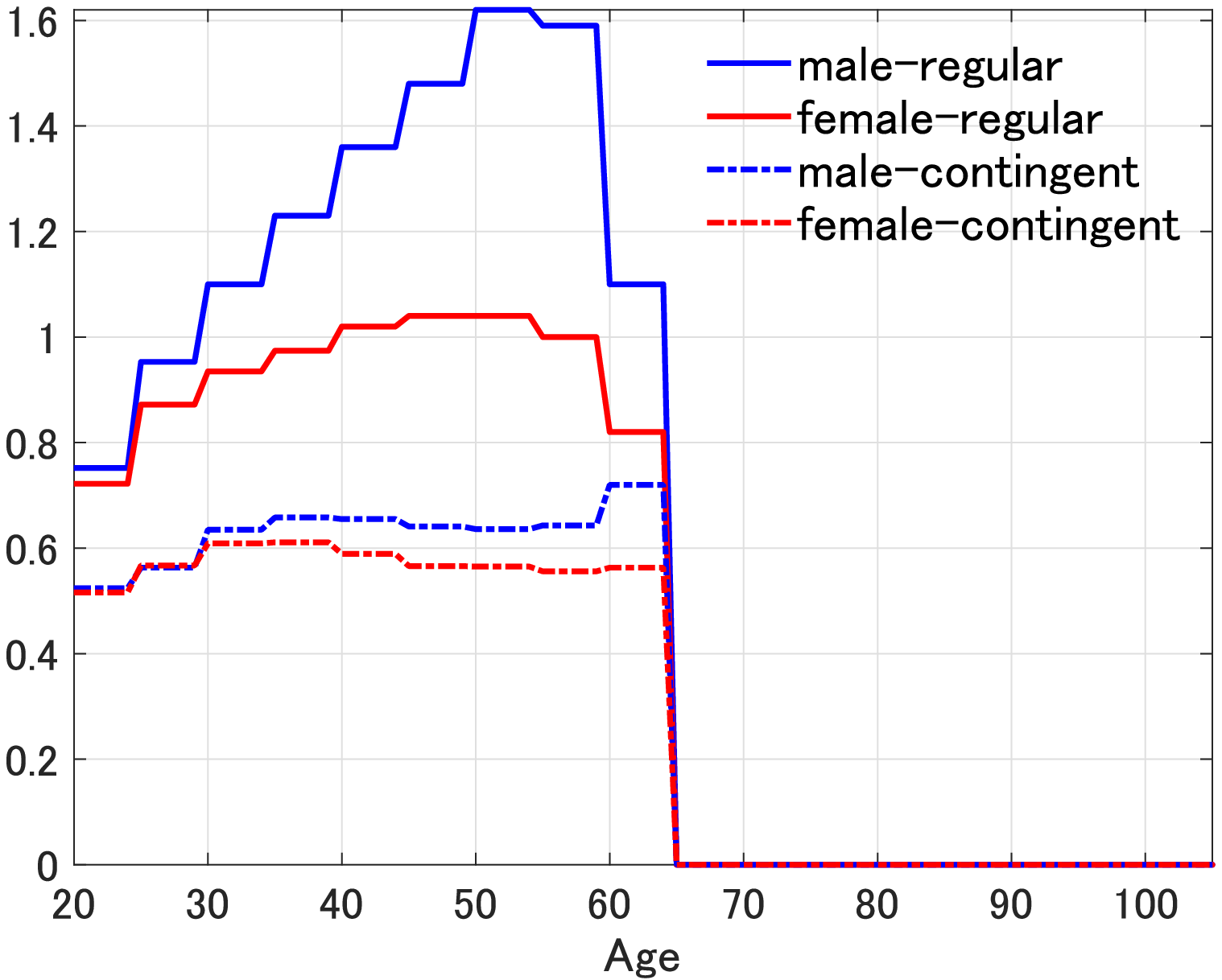}

(a)~ Survival rate \hspace{11em}
(b)~ Labor productivity
\end{center}
\end{minipage}

%\vspace{0.5\baselineskip}
\caption{Survival rate and labor productivity}
\label{fig:sr-eff}
%\vspace{0.5\baselineskip}

\end{figure}

\begin{figure}[tpb]
 \centering
% \vspace{-0.1\baselineskip}
 \includegraphics[width=130mm] %, height=77mm]
 {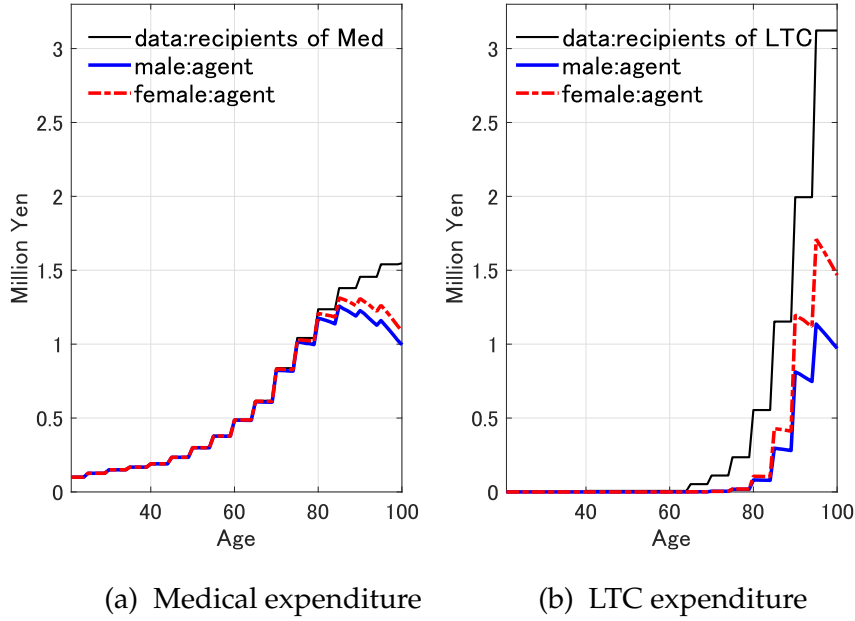}

 \hspace{1em}
 (a)~ Medical expenditure \hspace{3.5em}
 (b)~ LTC expenditure 

 \vspace{-0.2\baselineskip}
 \caption{Expected medical and LTC expenditure (per capita)}
 %\vspace{0.3\baselineskip}
 \label{fig:expected-M-Lexpend}
\end{figure}

%We suppose that households enter the economy at the age of 20, and they live up to a maximum age of 120.
We suppose that households enter the economy at 20 years and live up to 120 years.
In the notation of our model, this corresponds to age $j={1,2,\dots,100}$.
The four types of households are characterized by
(1) survival rates,
(2) labor productivity, and
(3) lifetime expected medical and LTC expenditures.
Figure \ref{fig:sr-eff} shows (1) and (2), and Figure \ref{fig:expected-M-Lexpend} shows (3). 
First, we suppose that agents are characterized by the age- and gender-specific survival rates $s_{j,g,t}$ and that the growth rate of a new cohort $n_{g,t}$ is based on the population distribution and survival rate projections of the IPSS (2017 estimates). 
\footnote{
The future population projections by the IPSS are formal estimates for 2015--2065 and are calculated by setting the mortality and fertility rates at that time will remain at the same rate after 2065.}
Next, to set up heterogeneous agents by dividing households into gender and employment type, the population is assigned to the four types based on the composition ratio of each type described in the Basic Survey on Wage Structure Statistics (2019) by the Ministry of Health, Labour and Welfare. 
%%%%追加220913
%なお、我々は家計のタイプは将来にわたって一定で推移する、つまり家計はライフサイクルを通じて雇用形態を変更しないと仮定する。
%注：この点は我々の分析の課題の1つであり、本来は雇用形態の選択は家計の重要な意思決定である。たとえば、yamaguchi (2019)やKM(2022)では、家計のミクロデータを用いて遷移行列を設定し、雇用形態の選択をモデルに組み入れている。我々はミクロデータを用いていないため、こうした個人の意思決定をモデルに組み入れていない。
%ただし、Yamaguchi (2019) などは、日本の女性のパネルデータを用いて前期から今期にかけて雇用形態を変化させる人の割合は大きくないことを示している。前期にregualr worker だった者が今期もregualr workerである確率は82\%程度である一方、contingent workerとなる確率は4\%程度である。前期にcontingent workerである場合の遷移確率も同様である。よって我々の分析では、雇用形態の選択が一定であると仮定している。
Furthermore, we suppose that this composition ratio is constant and that households do not change their employment status in the future.
\footnote{
This is one of the challenges of our analysis. The choice of employment type is an important decision made by households.
For instance, \citet{yamaguchiEffectsParentalLeave2019}
 and \citet{KSMM2022FLFP} use household microdata to set up transition matrices and incorporate the choice of employment status into their models. We do not use microdata and thus do not incorporate these individual decisions into our model.
However, \citet{yamaguchiEffectsParentalLeave2019}, for example, shows using panel data for Japanese women that the share of those who change their employment status from the previous period to the current period is small.Individuals who were regular workers in the previous period continue to be regular workers in the current period with a probability of about 82\%, while changing contingent workers in the current period with a probability of about 4\%. The transition probability of being a contingent worker in the previous period is also a similar trend. 
Therefore, our analysis supposes that the employment type choice is constant for simplicity.}

%%%%%%%%%%%%%%%%%%%%%%%%%%%%%%%%%%%%%%%%%%%%
%%%%%%%%%%%%%%%%%%%%%%%%%%%%%%%%%%%%%%%%%%%%% 

We use the per capita medical and LTC expenditures presented by \citet{IwamotoFukui2018}.
The expected medical expenditure $m_{j,t}^{H}$ is obtained by multiplying the gender-specific survival rates presented in Figure \ref{fig:sr-eff} Panel (a) and the per capita age-specific expenditure data shown in Figure \ref{fig:Med-LTC_expend}. 
Furthermore, the expected LTC expenditures $ m_{j,t}^{L}$ is obtained by multiplying the gender-specific survival rates, the certification rate of LTC need, and the per capita age-specific expenditure data mentioned above. 
Figure \ref{fig:expected-M-Lexpend} Panel (a) and (b) show the expected medical and LTC expenditures per capita, respectively.
\footnote{
We define the certification rate of long-term care needs as the number of persons certified needing LTC (\textit{you-kaigo}) level one to five, excluding persons certified needing support (\textit{you-shien}), divided by the population in each age bracket.
This ratio is higher the larger the number of persons certified needing LTC level one to five.}
Because the survival rates are higher for females than for males, especially for those aged 60 and older, the expected medical and LTC expenditures are also higher for females.

%%%%%%%%%%%%%%%%%%%%%%%%%%%%%%%%%%%%%%%%%%%%
%%%%%%%%%%%%%%%%%%%%%%%%%%%%%%%%%%%%%%%%%%%%% 

\subsection{Preferences}

\begin{figure}[ht]
 \centering

\vspace{0.5\baselineskip}
\noindent\begin{minipage}[t]{1\columnwidth}%
\begin{center}

\vspace{0.5\baselineskip}
\hspace{1em}
\includegraphics[width=67mm, height=57mm]{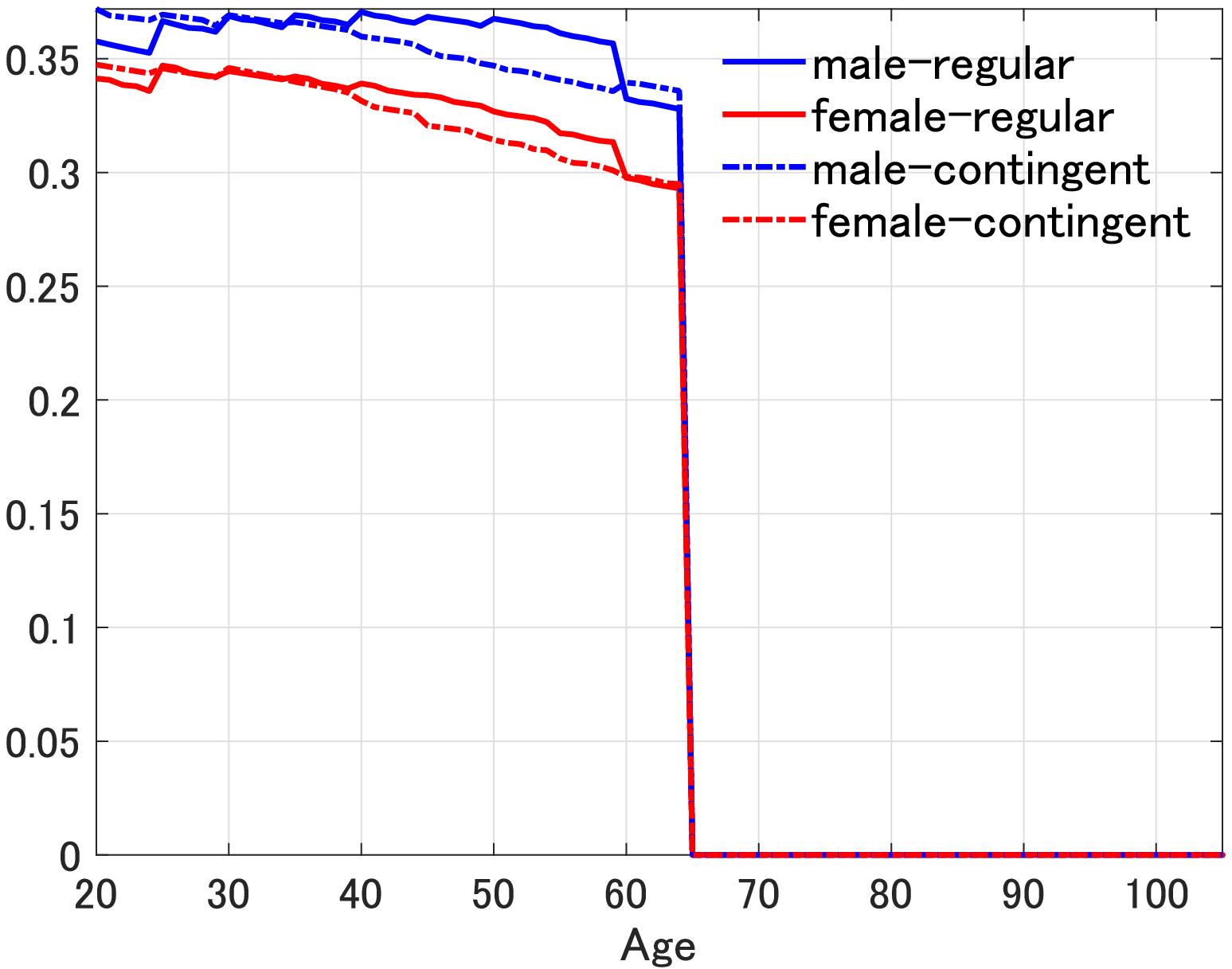}\hspace{3em}
\includegraphics[width=67mm, height=57mm]{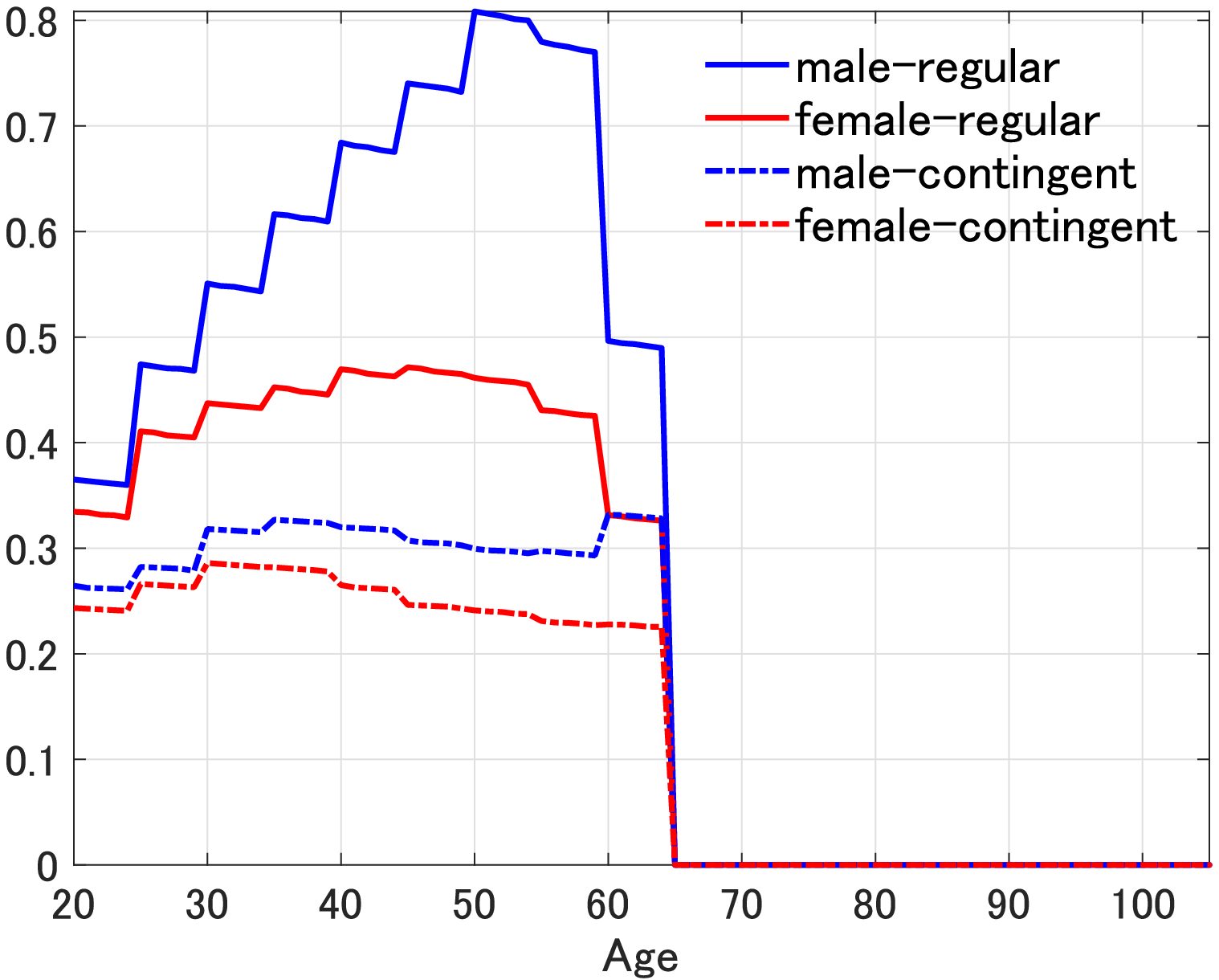}

~~
(a)~ Labor supply \hspace{10.7em}
(b)~ Labor income

\vspace{0.8\baselineskip}
\hspace{2em}
\includegraphics[width=62mm, height=57mm]{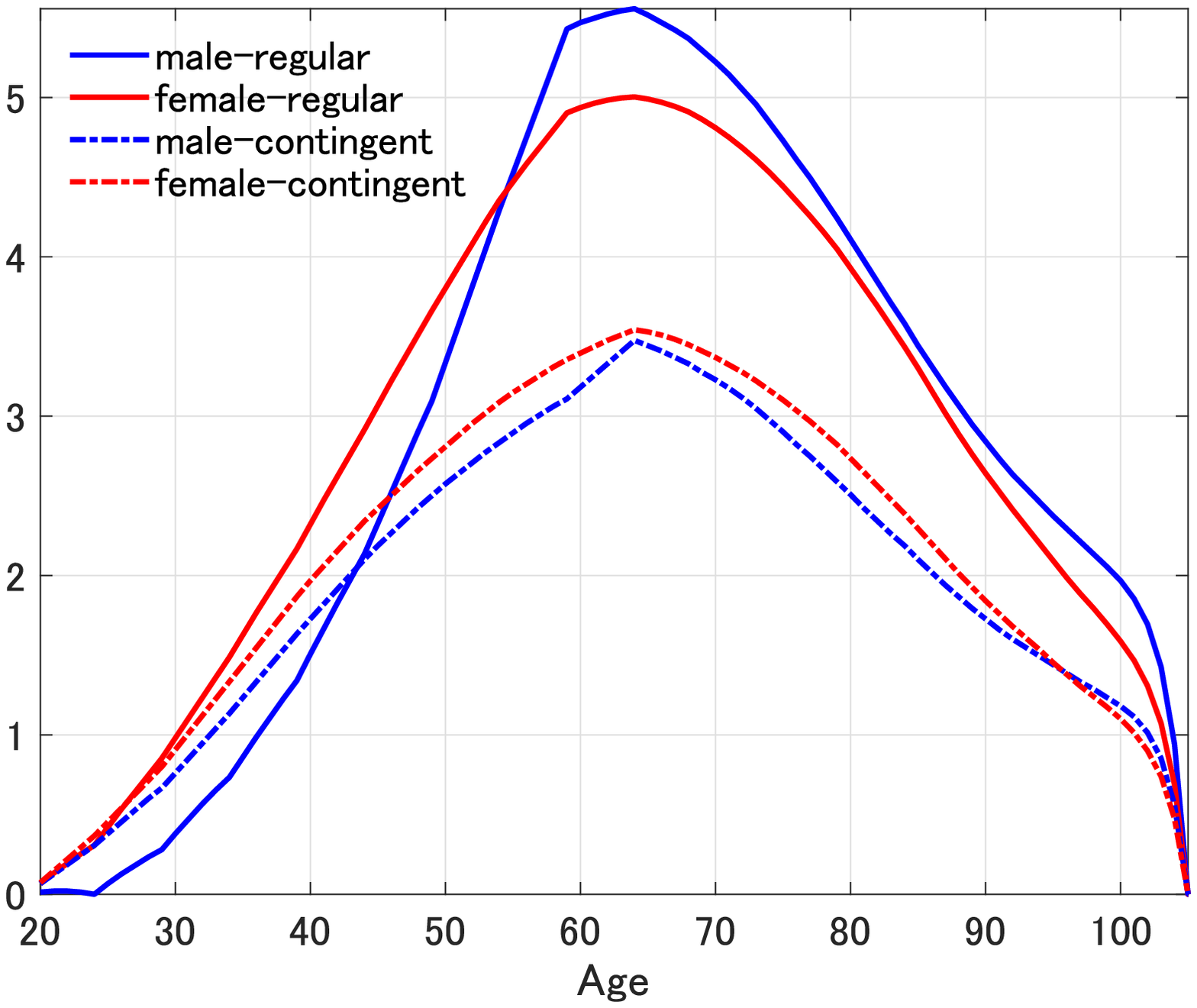}\hspace{2.2em}
\includegraphics[width=67mm, height=57mm]{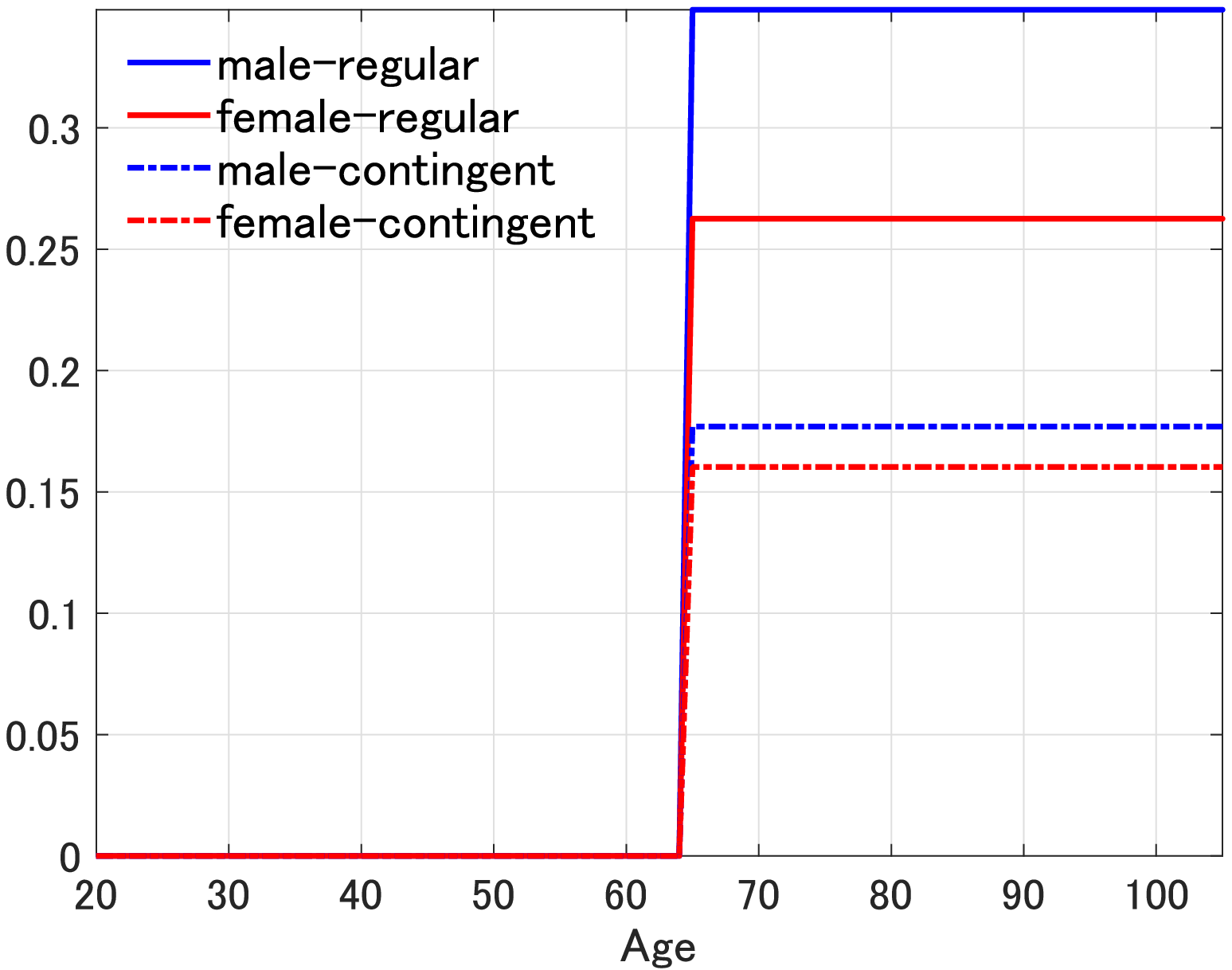}

~~~~~~
(c)~ Assets \hspace{15em}
(d)~ Pensions 

\end{center}

\end{minipage}
\caption{Distribution of individual variables in the initial steady state}
\label{fig:calculation_BGP}

\end{figure}

As shown in equation \eqref{eq:utility}, an agent's utility function is composed of consumption and labor supply, and, while the log function is used for consumption, the CES function is adopted for the disutility of labor.
\cite{KuroYama2008} estimated the value of Frisch labor supply elasticity $\zeta^{g}$ ($g \in \{m, f\}$; hereafter Frisch elasticity), which differs between males and females, using data in Japan.
Therefore, we conduct our analysis based on their estimation results.
The Frisch elasticity in gender units is set as $\zeta^{m} = 0.03$ for males, and $\zeta^{f} = 0.05$ for females, and the preference parameter of leisure $\gamma = 10$ for both males and females.
\begin{align}
u_{i}(c_{t},l_{t}) = \text{log}(c_{t}) + \gamma\,\frac{(1 - l_{t})^{1 - \zeta^{g}}}{1 - \zeta^{g}}.
\label{eq:utility}
\end{align}
The labor supply in the initial steady state of the four heterogeneous agents calculated using the parameters described above is shown in Figure \ref{fig:calculation_BGP} Panel (a).
As can be seen in this figure, regardless of the age, labor income, gender, or employment type of the worker, agents spent around eight hours a day (about one-third of a day) working. 
In addition, the labor supply is higher for males than for females.
These are consistent with the results of empirical studies such as the one \citet{KuroYama2008}.

%%%%%%%%%%%%%%%%%%%%%%%%%%%%%%%%%%%%%%%%%%%%
%%%%%%%%%%%%%%%%%%%%%%%%%%%%%%%%%%%%%%%%%%%%% 

\subsection{Technology}

The technology parameters in Table \ref{tab:parameters} govern the technology growth rate ($\gamma_A=0.3 \%$), investment rate, and capital income share in both sectors.
The long-run growth rate for labor-augmented technology $\gamma_\Omega$ is set at 0.7\%. 
By combining the two technology growth rates with the long-run population growth rate $\gamma_n$ indicating the demographic parameters as $-1.0$\%, then the long-run GDP growth rate is offset as just 0\%. 

In Table \ref{tab:parameters}, the share parameter of the aggregate production function, equation \eqref{eq:production_func} $\eta_1$ (which determines the relative share of income for Sector 1) is set to 64\%. This parameter is based on the share of operating surplus and mixed income in Sector 1.
Following \citet{MMP2018IMF},
%■McGrattan et al. (2019IMF), 
we use the JSNA data (average from 2015 to 2019) to determine the relative amounts of investment and fixed assets in the two sectors of production. 
Thus, by choosing the tangible capital ratio ($\theta_{1T}$, $\theta_{2T}$) and the tangible depreciation rate ($\delta_{1T}$, $\delta_{2T}$), we can ensure that the model's investment and fixed assets are consistent with the JSNA's tangible investment and equity (see table \ref{tab:assettable}). 
Accordingly, the tangible capital shares of the two sectors are calibrated as $\theta_{1T}=0.45$ and $\theta_{2T}=0.35$. 
The annual depreciation rates that produce investment rates consistent with the Japanese data are $\delta_{1T}=0.08$ and $\delta_{2T}=0.05$.
%%% capital ratioもSector 1の方が高いのでは？（22/05/08差し替え）
%The slightly high capital ratio and low depreciation rate in Sector 2 is due to inclusion of housing for the household business sector.
The tangible capital share and depreciation rate in Sector 2 include housing costs for the household business sector.
The intangible capital ratio and depreciation rates, $\theta_{1I}$, $\theta_{2I}$, $\delta_{1I}$, and $\delta_{2I}$, cannot be identified uniquely with the data that we have. In our baseline model, we calibrated these parameters following \citet{AratoYama2012}
%■Arato and Yama (2012) 
who derived estimates of tangible and intangible assets for Sector 1. 
To calibrate the intangible assets in Sector 2, we suppose the same ratio of intangible assets and non-land fixed assets. 
As a result, we set these parameters to
$\theta_{1I}=0.15$, 
$\theta_{2I}=0.05$, 
$\delta_{1I}=0.08$, and
$\delta_{2I}=0.09$.

%%%%%%%%%%%%%%%%%%%%%%%%%%%%%%%%%%%%%%%%%%%%%%%%%%%%5
\begin{table}[hbtp]
\begin{footnotesize}
\caption{Parameters of the model economy calibrated to Japan's data}
\label{tab:parameters}

\begin{center}
\begin{tabular}[t]{llc}
\hline 
Parameters & Descriptions & Values\tabularnewline
\hline 
$\gamma^{n}$ & Growth rate of the population (long-run) & $-1\%$ \tabularnewline
$J^{R}$ & Retirement age & 46 \tabularnewline
$J$ & Maximum possible age & 101 \tabularnewline
%%% Number of workers per retiree = 3.93はMcgrattan & Prescott (2017)とまったく同じで大丈夫か？＆本文で言及無いのでとりあえず削除。
% & Number of workers per retiree & 3.93 \tabularnewline
% &  & \tabularnewline
\hline 
\multicolumn{2}{l}{\textbf{Preference parameters}} & \tabularnewline
$\gamma$ & Preference parameter of leisure & 10.0\tabularnewline
$\zeta^{m}$ & Frisch elasticity for male & 0.03\tabularnewline
$\zeta^{f}$ & Frisch elasticity for female & 0.05\tabularnewline
$\beta$ & Discount factor & 0.983 \tabularnewline
% &  & \tabularnewline
\hline 
\multicolumn{2}{l}{\textbf{Technology parameters} } & \tabularnewline
$\gamma^{A}$ & TFP growth rate & 0.3\% \tabularnewline
$\gamma^{\Omega}$ & Labor-augmented productivity growth rate & 0.7\%\tabularnewline
$\eta_{1}$ & Income share of corporate sector & 0.640 \tabularnewline
% &  & \tabularnewline
\hline 
\multicolumn{2}{l}{\textbf{Interest rate of bond}} & \tabularnewline
$i_t^d$ & Interest rate on government bonds & 0.01 \tabularnewline
\hline 
\multicolumn{2}{l}{\textbf{Capital shares} } & \tabularnewline
$\theta_{1T}$ & Tangible capital of corporate sector & 0.450 \tabularnewline
$\theta_{1I}$ & Intangible capital of corporate sector & 0.150 \tabularnewline
$\theta_{2T}$ & Tangible capital of household business & 0.350 \tabularnewline
$\theta_{2I}$ & Intangible capital of household business & 0.050 \tabularnewline
% &  & \tabularnewline
\hline 
\multicolumn{2}{l}{\textbf{Depreciation rates }} & \tabularnewline
$\delta_{1T}$ & Tangible capital of corporate sector  & 0.080 \tabularnewline
$\delta_{1I}$ & Intangible capital of corporate sector & 0.080 \tabularnewline
$\delta_{2T}$ & Tangible capital of household business & 0.050 \tabularnewline
$\delta_{2I}$ & Intangible capital of household business & 0.090\tabularnewline
% &  & \tabularnewline
\hline 
\multicolumn{2}{l}{\textbf{Fiscal policy parameters}} & \tabularnewline
$\psi_{G}$ & Government consumption & 0.190 \tabularnewline
$\psi_{B}$ & Government debt & 1.500 \tabularnewline
% &  & \tabularnewline
\hline 
\multicolumn{2}{l}{\textbf{Capital tax rates}} & \tabularnewline
$\tau_{1}^{\pi}$ & Corporate income tax & 0.250\tabularnewline
$\tau_{1}^{d}$ & Tax on corporate distributions & 0.250 \tabularnewline
$\tau_{2}^{d}$ & Tax on household business distributions & 0.250 \tabularnewline
\hline 
\end{tabular}
\end{center}

\begin{enumerate}
\item[Notes:] The set values of parameters other than the preference parameters rely on the values from \citet{MMP2018IMF}.
The values of the labor supply parameters are derived from the estimation of the labor supply elasticity of  \citet{KuroYama2008}.
%The technology parameters in Table \ref{tab:parameters} govern technological growth, investment rates, and capital income shares across business sectors. The long run growth rate of TFP is set equal to 0.3 \%. Together with the long term population growth rate of $-1$ \% as implied by the demographic parameters. The share parameter in the aggregate production function, which determines the relative share of income to private corporations, is set equal to 64 \%. This parameter is based on the private corporates' proportion of operating surplus and mixed income.
\end{enumerate}
\end{footnotesize}
\end{table}
%%%%%%%%%%%%%%%%%%%%%%%%%%%%%%%%%%%%%%%%%%%5

\subsection{Employment, labor income, and assets}

In our model, $e^{j,g,h}$ represents the efficiency units supplied to the labor market by a group of agents of age $j$, gender $g$, and employment type $h$. 
To establish the labor productivity of the four types, that is, combinations of male or female and regular workers or contingent workers, we calculate the hourly wages for each type of worker. 
This result is shown in Figure \ref{fig:sr-eff} Panel (b).
For data on wages and working hours, we use the Basic Survey on Wage Structure Statistics (2019).
Using these data, we calculate hourly wages for each male and female, regular and contingent worker, and then we estimate labor productivity based on the method presented by \citet{Hansen1993} 
and used by 
\citet{BIJ2006},
\citet{Yamada2011},
and others. 
Figure \ref{fig:calculation_BGP} Panel (b) shows the age distribution of labor income calculated in the initial steady state based on this labor productivity.
\footnote{
\citet{KSMM2020JJIE} used not only the Basic Survey on Wage Structure (BSWS) but also the Employment Status Survey by the Statistics Bureau of the Ministry of Internal Affairs and Communications to determine the productivity of the self-employed, whereas, in our paper, the four types of productivity described above are determined using solely the data from the BSWS.}
In our analysis, since the baseline retirement age is set uniformly at 65, labor productivity is set so that it will reach zero at age 65. 
For both male and female regular workers, the hourly wages and labor productivity decline sharply after reaching the age of 60, but this is thought to reflect the decline in wages due to reemployment after the retirement age of 60, which is actually adopted by most companies.

The distribution of assets for each agent in the initial steady state is shown in Figure \ref{fig:calculation_BGP} Panel (c). 
This distribution indicates that male regular workers accumulate fewer assets at a younger age than female regular workers and that they begin to increase them rapidly after the age of 40.
After the age of 50, the assets of male regular workers are lager than those of female regular workers.
The reason for this lies in the shape of the labor productivity curve shown in Figure \ref{fig:sr-eff} (b) and the labor income shown in Figure \ref{fig:calculation_BGP} Panel (b).
That is, the productivity and income of male regular workers increase significantly by the peak in their early 50s compared with female regular workers.
Male regular workers are aware of this and thus are able to allocate a larger share of their income to consumption in their younger years.

\subsection{Government}

\paragraph{Social security system and fiscal balance}
The government runs a PAYG pension system. 
In the baseline simulation, the normal retirement age $J^R$, equivalent to age 65, is set at 46. 
As shown in Figure \ref{fig:calculation_BGP} Panel (d), pension payments are calculated from an income replacement rate of 62\%.
Medical and LTC costs per capita are based on the amounts published in the Basic Data on Medical Insurance and the Statistics of Long-Term Care Benefit Expenditures by the Ministry of Health, Labour and Welfare, and are calculated following \citet{IwamotoFukui2018}.
In our quantitative analysis, we set the upper limit of the government debt-to-GDP ratio at 1.5 in the steady state. 
In addition, we suppose that the fiscal balance is adjusted by changes in the consumption tax during the transition path.

\paragraph{Tax system}
Progressive taxation is set according to the income bracket of the household. 
\citet{MMP2018IMF} use income and promotion tax tables estimated by IMF staff using National Tax Agency data. 
In our study, we also use the tax rates per income bracket they estimated. The details are shown in Table \ref{tab:taxtable}.
Here, we suppose that the progressive labor tax of our model consists mainly of national and local taxes, such as inhabitant taxes, public pension, public medical insurance, and LTC insurance premiums.

%%%%%%%%%%%%%%%%%%%%%%%%%%%%%%
\begin{table}[t]%[hbtp]
\begin{footnotesize}
\caption{Progressive tax table for labor tax}
\label{tab:taxtable}

\begin{center}
\begin{tabular}[t]{l|c|c|c|c}
\hline 
Labor earnings  & Share  & Tax rate  & $\alpha$  & $\beta$ \tabularnewline
\hline 
Under 1  & 17.3  & 0.027  & 0  & 0 \tabularnewline
1--2  & 15.0  & 0.191  & $-0.0164$  & 0.0027 \tabularnewline
2--3  & 15.5  & 0.272  & $-0.0326$  & 0.0218 \tabularnewline
3--4  & 15.4  & 0.285  & $-0.0365$  & 0.049 \tabularnewline
4--5  & 12.2  & 0.294  & $-0.0401$  & 0.0775 \tabularnewline
5--6  & 8.3  & 0.302  & $-0.0441$  & 0.1069 \tabularnewline
6--7  & 5.1  & 0.306  & $-0.0465$  & 0.1371 \tabularnewline
7--8  & 3.5  & 0.315  & $-0.0528$  & 0.1677 \tabularnewline
8--9  & 2.4  & 0.324  & $-0.0600$  & 0.1992 \tabularnewline
9--10  & 1.5  & 0.328  & $-0.0636$  & 0.2316 \tabularnewline
10--15  & 2.8  & 0.338  & $-0.0736$  & 0.2644 \tabularnewline
15--20  & 0.6  & 0.358  & $-0.1036$  & 0.4334 \tabularnewline
20--25  & 0.2  & 0.387  & $-0.1616$  & 0.6124 \tabularnewline
Over 25  & 0.2  & 0.447  & $-0.3116$  & 0.8059 \tabularnewline
\hline 
Total  & 100.0  &  &  & \tabularnewline
\hline 
\end{tabular}
\end{center}

\begin{enumerate}
\item [Notes:] Following \citet{MMP2018IMF}, the second column "Share" is obtained from the distribution of labor income based on the Tax Surveys published by the National Tax Agency. 
The third column "Tax rate" also contains the estimates of effective income tax rates across 14 income brackets obtained from \citet{MMP2018IMF}. 
The fourth and fifth columns "$\alpha$" and "$\beta$", which are calculated from the tax rate, are referred to as the intercept and slope of the progressive tax structure, respectively. 
\end{enumerate}
\end{footnotesize}
\end{table}
%%%%%%%%%%%%%%%%%%%%%%%%%%%%%%%%%%%%%%%%%%%%%%

The capital tax rates are shown in Table \ref{tab:parameters}.
The effective corporate income tax rate $\tau_{1}^{\pi}$ is set at 25\% to bring the modeled values in line with the corporate income tax and other corporate taxes paid by corporations in the JSNA. 
In addition, investors in corporations in which distributions are paid in the form of dividends will pay an additional tax on distributions. 
These tax rates $\tau_{1}^{d}, \tau_{2}^{d}$ are set at 25 \% to match the capital income and the size of the corporate sector.
\footnote{
These tax rates are set following \citet{MMP2018IMF}.}

Table \ref{tab:parameters} also contains the fiscal policy parameters.
$\psi_B$ is defined as the total debt of the general government minus the financial assets held by public pension funds. 
These assets are accumulated for future pension liabilities, and we also incorporate them into our model. 
The level of government consumption $\psi_G$ is set to a constant percentage of adjusted GDP over the entire period.

%%%%%%%%%%%%%%%%%%%%%%%%%%%%%%%%%%%%%%%%%%%%%%%%%%%%%%%%%%%%%%%%%%%%%%%%%%%%%%%%%%%%
\section{Numerical analysis}\label{sec:NA}
%%%%%%%%%%%%%%%%%%%%%%%
% ★memo★シナリオ：労働力人口や経済環境を変化させ、影響が長期的に続く。
% (1)ベースライン
% (2)定年延長

%★政策：社会保障改革。短期的な影響・ずっと将来には影響を与えなくなる
% (1) Current Policy　→現行政策の継続
% (2) 年金代替率引き下げ（2015年時点で62%から2047年に50.8%へ線形に引き下げる。2019年財政検証の経済前提ケースⅢを想定）
% (3) 高齢者の医療自己負担を10％up（70歳未満は3割負担、70歳以上75歳未満は2割負担→3割負担、75歳以上は1割負担→2割負担）2030年に
% (4) 介護自己負担を10%up（40歳以上1割負担→2割負担に）2030年に
%%%%%%%%%%%%%%%%%%%%%%%

%このsectionでは、移行過程の定量分析を通じた政策シミュレーションの結果について述べる。
%我々は、「ベースライン」と「定年延長」という2つのシナリオに基づいて分析する。
%この2つのシナリオごとに、以下で述べる3つの社会保障改革それぞれについて政策シミュレーションを行う。
%そして、性別と雇用形態で区別された4タイプの家計ごとに現在の勤労世代、引退世代、および将来世代への影響を示し、それぞれの影響の違いを検証する。
In this section, we describe the quantitative results of a policy simulation for the transition path.
Our simulation is based on two scenarios: (i) the \textit{baseline} and (ii) the \textit{extension of the retirement age}.
For these two scenarios, we conduct policy simulations for the three options of social security reforms described below.
Then, we calculate the impact on the current working, retired, and future generations for the four types of households distinguished by gender and employment type, and evaluate the extent to which differences in the impact are generated for the four types.

%本節では、次の3つのステップで議論する。
%第1に、2つのシナリオで想定する経済状態と、3つの社会保障改革の概要を述べる。
%第2に、2つのシナリオのもとで社会保障改革が実施された場合に、主要なaggregateの経済変数のtransition pathがどのように変化するかを示す。
%特に、実質GDPの推移と消費税率の水準の推移に着目する。
%第3に、社会保障改革が人々の厚生に与える影響を、現在の勤労世代および引退世代と将来世代ごとに計算する。
%そして、家計のタイプに基づく異質性に着目して、改革の効果について議論する。
We proceed with the following three aspects.
First, we outline the economic circumstances  underlying  the two scenarios and the three options of social security reforms.
Second, we show how the transition paths of key aggregate economic variables would change if one of the three reforms is implemented in the two scenarios.
In particular, we focus on changes in the real GDP and the level of the consumption tax rate.
Third, we compare the impacts of the different reforms on the welfare for the current working, retired, and future generations.
We then discuss the effects of the reforms, focusing on the inter-generation and intra-generation heterogeneity of households.

\subsection{Scenarios and social security reforms}%5.1項
%このsubsectionでは、我々の分析における2つのシナリオと3つの社会保障改革の内容について述べる。
%%%■これは止めた■\footnote{AppendixではTFP高成長のシナリオについても考察する。}
%Section \ref{sec:CALIB}で述べた通り、我々は(1) ベースライン・シナリオにおいてTFP成長率γ^Aは0.3\%、定年年齢J^Rは65歳、長期での人口成長率γ^nは-1\%と仮定する。
%(2) 定年延長シナリオでは、2030年に定年年齢J^Rが65歳から70歳に上昇するという点のみ、ベースラインから変更する。
Here, we describe the two scenarios and three social security reforms in our analysis.
%%%■これはやめた◆¥footnote{We also discusses the TFP high growth scenario in the Appendix.}
As mentioned in section \ref{sec:CALIB}, we suppose that 
(i) the TFP growth rate $\gamma_A$ is 0.3\%, 
the long-run labor-augmented technology growth rate $\gamma_\Omega$ is 0.7\%,
the retirement age $J^R$ is 46 (the age of 65), and the long-run population growth rate $\gamma_n$ is $-1$\% in the baseline scenario.
In the extended retirement age scenario, (ii) the only change from the baseline is that the retirement age $J^R$ increases from 65 to 70 in 2030.

%この2つのシナリオにおいて、我々は、社会保障改革を何も行わなかった場合と、次の3つの改革が実施された場合の政策シミュレーションを行う。
%すなわち、(1)現行政策の維持、(2)年金所得代替率κ_tの段階的な引き下げ、(3)医療支出の自己負担率λ_(j,t)^hの引き上げ、(4)介護支出の自己負担率λ_(j,t)^lの引き上げ、である。
%政策(1) では、年金所得代替率κ_tは62\%のまま維持され、医療支出の自己負担率λ_(j,t)^hは20歳以上70歳未満の人々は30\%、70歳以上75歳未満は20\%負担、75歳以上は10\%負担、介護支出の自己負担率λ_(j,t)^lは全世代で10\%負担が想定される。
%政策(2) では、section \ref{sec:INTRO} で言及したように、2015年時点の62%から2047年に50.8%まで線形に引き下げていく改革を想定する。
%■\footnote{この想定は、厚生労働省「2019年将来の公的年金の財政見通し」におけるケースIIIに基づいている（https://www.mhlw.go.jp/stf/seisakunitsuite/bunya/nenkin/nenkin/zaisei-kensyo/index.html　2021年3月2日アクセス）。}
%政策(3) では、2030年に高齢者の医療支出の自己負担率を引き上げて、全年齢で一律30%とする改革を想定する。 
%%◆これはカット■\footnote{Section \%ref{sec:INTRO}でも述べたように、2022年後半から75歳以上の医療費の自己負担率が2割に引き上げられることが決まっているが、本稿の議論はそれよりもさらに医療費の自己負担率を高めた想定となっている。}
%政策(4)では、2030年に介護支出の自己負担率を一律30\%とする改革を想定する。
In these two scenarios, we conduct policy simulations in the cases of the three reforms and the cases in which reforms are not implemented.
That is, 
(1) continuation of the current policy, 
(2) a gradual decrease of the pension income replacement rate $\kappa_t$, 
(3) an increase in the copayment rate $\lambda_{j,t}^{H}$ for medical expenditures, and
(4) an increase in the copayment rate $\lambda_{j,t}^{L}$ for LTC expenditures.
In policy (1), we suppose that the pension income replacement rate $\kappa_t$ is constant at 62\% and the copayment rate for medical expenditures $\lambda_{j,t}^{H}$ is 30\% for those aged 20 to 70, 20\% for those aged 70 to 75, 10\% for those aged 75 and over, and 10\% for LTC expenditures $\lambda_{j,t}^{L}$.
Policy (2) is a reform to reduce the pension income replacement rate from 62\% as of 2015 to 50.8\% in 2047.
\footnote{This assumption is based on Case III in the Ministry of Health, Labour and Welfare, "2019 Financial Projection of the Public Pension (Fiscal Verification)" (\url{https://www.mhlw.go.jp/stf/seisakunitsuite/bunya/nenkin/nenkin/zaisei-kensyo/index.} \\ \url{html}\quad accessed March 2, 2021).}
Policy (3) is a reform to raise the copayment rate for health care expenditures for the elderly in 2030 to a uniform 30\% for all ages. 
Policy (4) is a reform that would set the copayment rate for LTC expenditures at 30\%  in 2030.

\subsection{Transition of aggregate variables}%5.2項
%このsubsectionでは、実質GDPとその成長率、および消費税率のtransition pathの計算結果を示す。
%我々の分析では、2115年までは国立社会保障・人口問題研究所が公表する「日本の将来推計人口（2017年推計）」に基づいた人口成長率を用い、それ以降はどのエージェントも毎年-1\%と仮定する。
%なお、フルタイマー、パートタイマーの比率は2018年時点の賃金構造統計基本調査に基づいて定め、他の年はその比率で一定とする。
Here, we present the simulation results for the transition path of the real GDP, its growth rates, and the consumption tax rates under the demographic projection in which the population growth rates that are adopted are those of the IPSS (2017) up to 2115, and they are reduced by 1 \% per year after 2115.
The proportions of regular workers and contingent workers are supposed to be constant on the Basic Survey of Wage Structure Statistics in 2019 over the future.

\subsubsection{Transition of the output}
\begin{figure}[tp]
 \centering

 \includegraphics[width=165mm]{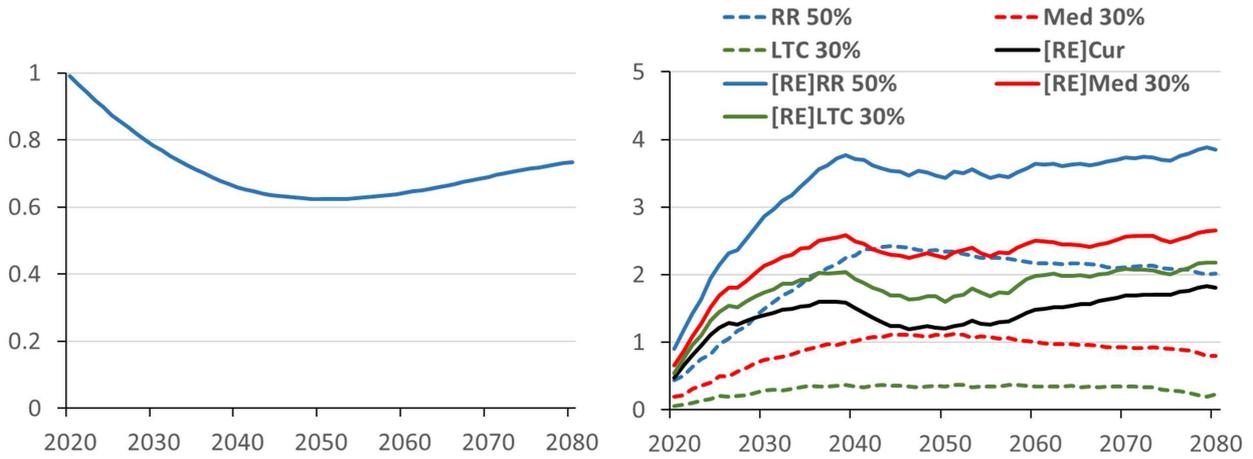}
 \vspace{-0.5\baselineskip}

 \hspace{4em}
 (a)~ Transition of the GDP \hspace{3em}
 \small{(b)~ Percentage change of the GDP induced by each reform}
 
 \vspace{-0.3\baselineskip}
 \caption{Transition of the real GDP and percentage change induced by each reforms}
 \label{fig:GDPtran}

 \begin{footnotesize}
 \begin{enumerate}
    \item[Notes:] [RE] indicates the extended retirement scenario.
 \end{enumerate}
 \end{footnotesize}
\end{figure}

%まず、outputのtransition pathに着目する。
%図\ref{fig:GDPtran}の左側のグラフは、ベースライン・シナリオにおける2020年の値を1に基準化したreal GDPの水準の推移が示されている。
%GDPは最も高齢化が進む2050年代半ばまではマイナス成長となっているが、それ以降はゆるやかに上昇しているのがわかる。
%\footnote{■Imrohorogru, Kitao and Yamada (2017guest worker論文, fig.2) でもoutputの長期的な推計が示されている。ここでは、Baselineで2014年から最初の20年間くらいはほぼ横ばいで推移し、その後急激に減少する結果が示されている。この点は、我々の結果とはやや異なる傾向である。このことは、我々のモデルでは無形資産が考慮されているため、GDPに対する資本蓄積の貢献が相対的に強く反映されることから、人口減少のペースが弱まる将来時点でGDPが下げ止まり、上昇に転じると考えられる。}
First, we focus on the transition path of the output.
Figure \ref{fig:GDPtran} Panel (a) shows the transition of the level of the real GDP, normalized to unity for the year 2020 in the baseline scenario.
Although GDP growth is downward until the mid-2050s, when the aging of the population will be at its peak, we see a gradual increase in GDP growth after around 2050.
\footnote{
\citet[fig. 2]{IKY2017EI} also present a long-term output estimate. They show a result in the baseline that remains almost flat for the first approximately 20 years starting in 2014 and then declines sharply. This is a somewhat different transition from our results. The reason for this is that our model includes intangible assets and thus reflects the relatively greater contribution of capital accumulation to GDP, which should lead to an increase in GDP at a future point in time when the pace of population decline weakens.}

%図\ref{fig:GDPtran}の右側のグラフでは、ベースライン・シナリオのreal GDPからのシナリオごとの\%変化が示されている。
%7つのすべてのシナリオで、ベースライン・シナリオと比べてGDPの水準が高く推移することがわかる。
%ここで、図のRR 50\%は年金代替率引き下げ、Med 30\%は医療支出の自己負担引き下げ、LTC 30\%は介護支出の自己負担引き下げ政策を示している。
%また、[RE]は定年延長シナリオを、何もない場合はベースライン・シナリオを示す。
%ベースラインにおける医療・介護支出の自己負担増の政策(赤色の破線、緑色の破線)は高々1%増程度である。
%一方、年金所得代替率引き下げ改革の場合は2.5\%程度まで増加する (青色の破線)。
%定年延長シナリオの場合は、政策間の差がより大きくなる。
%青色の実線は、年金所得代替率引き下げ改革の場合は4%弱まで増加していることを示している。
%この結果から、我々のモデルでは、定年延長したうえで段階的に年金所得代替率を引き下げる政策をとった場合に、GDPの将来に向けた推移の水準が最も高くなることがわかる。
Figure \ref{fig:GDPtran} Panel (b) shows the values of the comparisons by percentages of the level of the real GDP for each reform of both scenarios relative to the continuation of the current policy of the baseline scenario.
As shown in the legend in the graph, the blue, red, or green dashed line represents the reduction in the pension replacement rate (RR 50\%), a decreasing copayment for medical expenditures (Med 30\%), or a decreasing  copayment for LTC expenditures (LTC 30\%), in the baseline scenario, respectively. Similarly the blue, red, or green solid line represents RR 50\%, Med 30\%, or LTC 30\% in the extended retirement scenario (\hspace{0.1em}\small{[RE]}\hspace{0.1em}).
In all cases of reforms in the two scenarios, the level of GDP is higher than the case of no reform in the baseline scenario for all periods.
The reform of increasing copayments for medical and LTC expenditures in the baseline (the red dashed line and green dashed line) is at most a 1\% increase.
However, it increases to about 2.5\% in the case of reducing the pension income replacement rate (the blue dashed line).

In the extended retirement scenario, the difference between the policies is even larger.
The blue solid line shows an increase to just under 4\% in the case of a reduction of the pension income replacement rate.
This result shows that the level of GDP transition in the future is highest when the retirement age is extended and the reduction of the pension income replacement rate is introduced.

%%%%%%%%%%%%%%%%%%%%%%%%%%%%%%%%%%%%%%%%%%%%
%%%%%%%%%%%%%%%%%%%%%%%%%%%%%%%%%%%%%%%%%%%%% 

\begin{figure}[H]
 \centering

 \includegraphics[width=130mm]{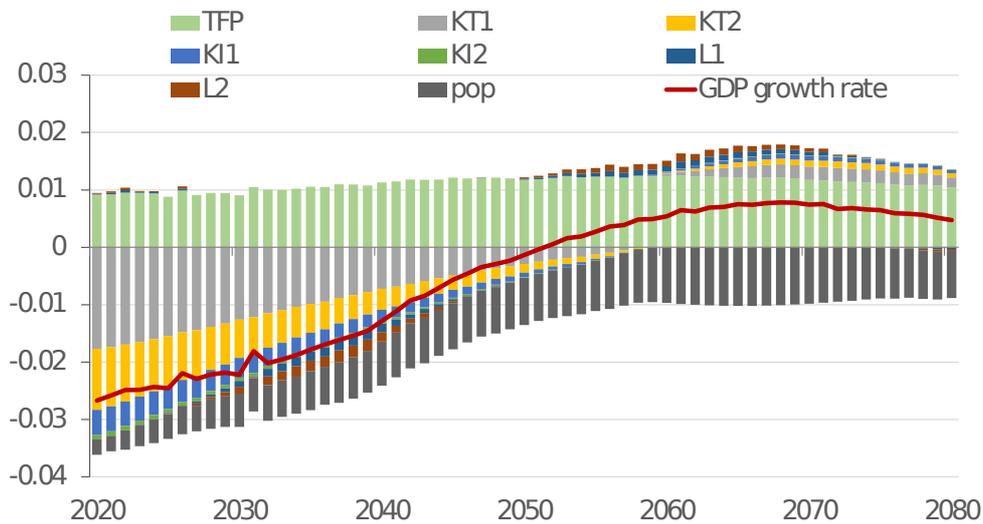}

 \vspace{-0.3\baselineskip}
 \caption{Output growth rate and its contribution decomposition}
 \label{fig:GDPtran_contribution}
\end{figure}

%図\ref{fig:GDPtran_contribution}は、ベースラインにおける無形資産も含めたoutputの成長率と、その寄与度分解の結果の推移を示している。
%section \ref{sec:MODEL}で述べた生産関数に基づいて、outputの成長率をTFP A_t、部門iの有形資産K_iT、無形資産有形資産K_iI、部門iの労働力L_i、および人口の成長率に分解している。
Figure \ref{fig:GDPtran_contribution} shows the growth rates of output in the baseline scenario, and the contribution decomposition of factors including intangible assets.
Based on the production function described in section \ref{sec:MODEL}, the growth rate of output is decomposed into TFP $A_t$, tangible assets $K_{iT}$ in sector $i$, intangible assets  $K_{iI}$, the labor force $L_i$ in sector $i$, and the population growth.

%%%%%%%%%%%%%%%%%%%%%%%%%%%%%%%%%%%%%%%%%%%%
%%%%%%%%%%%%%%%%%%%%%%%%%%%%%%%%%%%%%%%%%%%%% 

%TFPと労働生産性の合算の成長率(TFP、薄い緑色のバー)は仮定より安定的に約1\%の寄与度であるが、対照的に、人口の成長率(pop、濃いグレーのバー)は-1\%の寄与度である。
%この２つの寄与度はおおむねバランスしていることがわかる。
%他方で、2050年までoutputの成長率を下落させている大きな要因が、Sector 1の有形資産(KT1、薄いグレーのバー)と無形資産(KI1、黄色のバー)の成長率である。
%ここには、現役世代の減少と引退世代の増加は、貯蓄つまり資本の減少速度を加速させることが示されている。
%我々のシミュレーションは、output成長率に対する有形・無形資産の成長率（KT、KI）の寄与度は、労働力の減少(L1、L2)による寄与度よりもはるかに大きいという結果示している。
The total of the TFP growth rate and the labor-augmented technology growth rate (the light green bars) is stable at about 1\% for the contribution, following our setting.
In contrast, the growth rate of the population (dark gray bars) is shown as $-1$\% for the contribution. These two contributions are approximately offset to 0\% growth.
On the other hand, a major factor depressing the growth rate of output through 2050 is the growth rate of tangible assets in Sector 1 (light gray bars) and tangible assets in Sector 2 (yellow bars).

This is due to both a decrease in the labor force population and an increase in the retired population making the aggregate saving rate, or capital accumulation drop rapidly in this period.
Our simulations show that the size of the negative contribution of tangible and intangible assets to the output is much larger than that of the negative contribution of the labor force shown by the blue and brown bars.

\subsubsection{Transition of the consumption tax rate}
\begin{figure}[tp]
\centering

\noindent\begin{minipage}[t]{1\columnwidth}%
\begin{center}

%\vspace{0.2\baselineskip}
\hspace{2em}
\includegraphics[width=68mm]{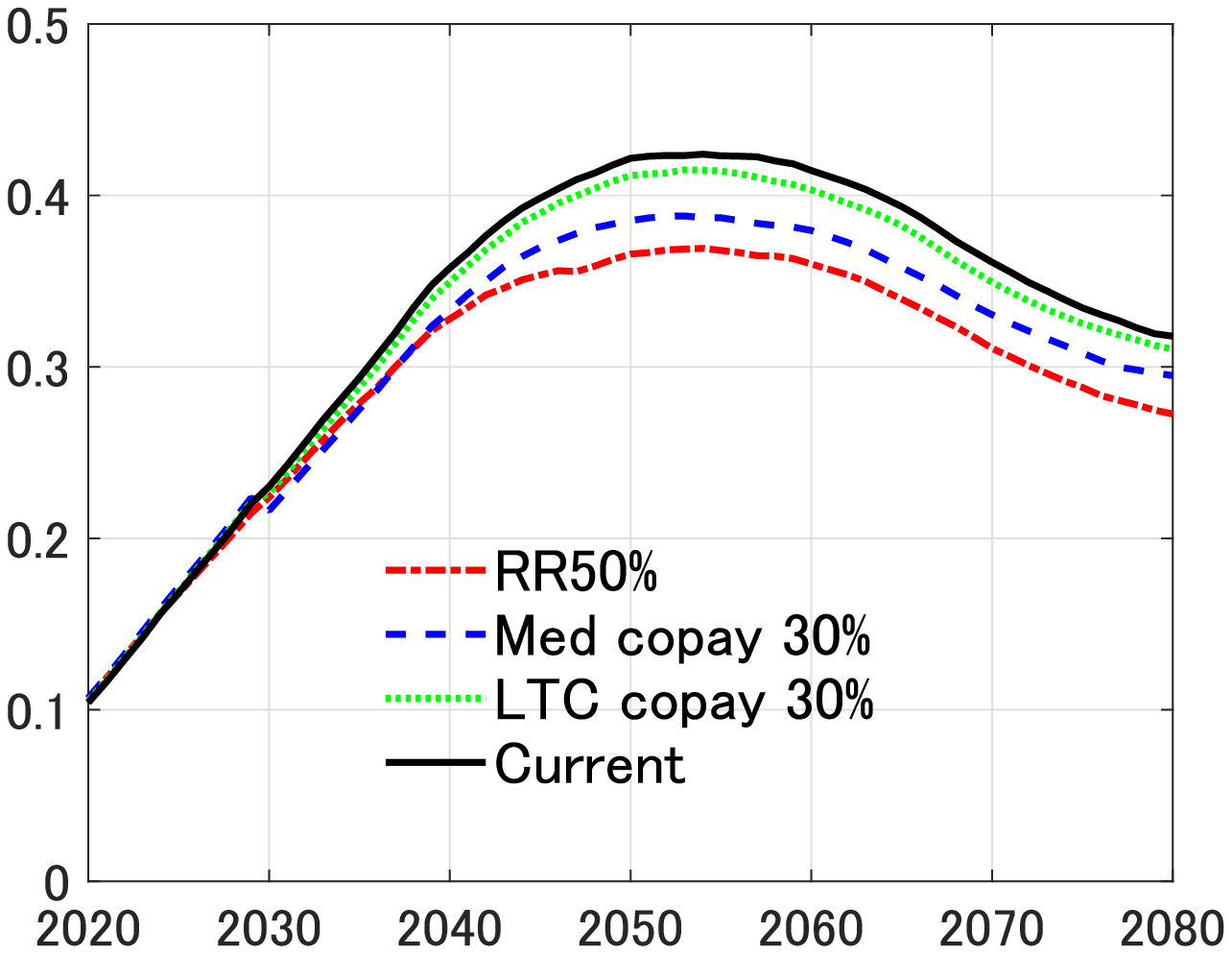} \hspace{1.5em}
\includegraphics[width=68mm]{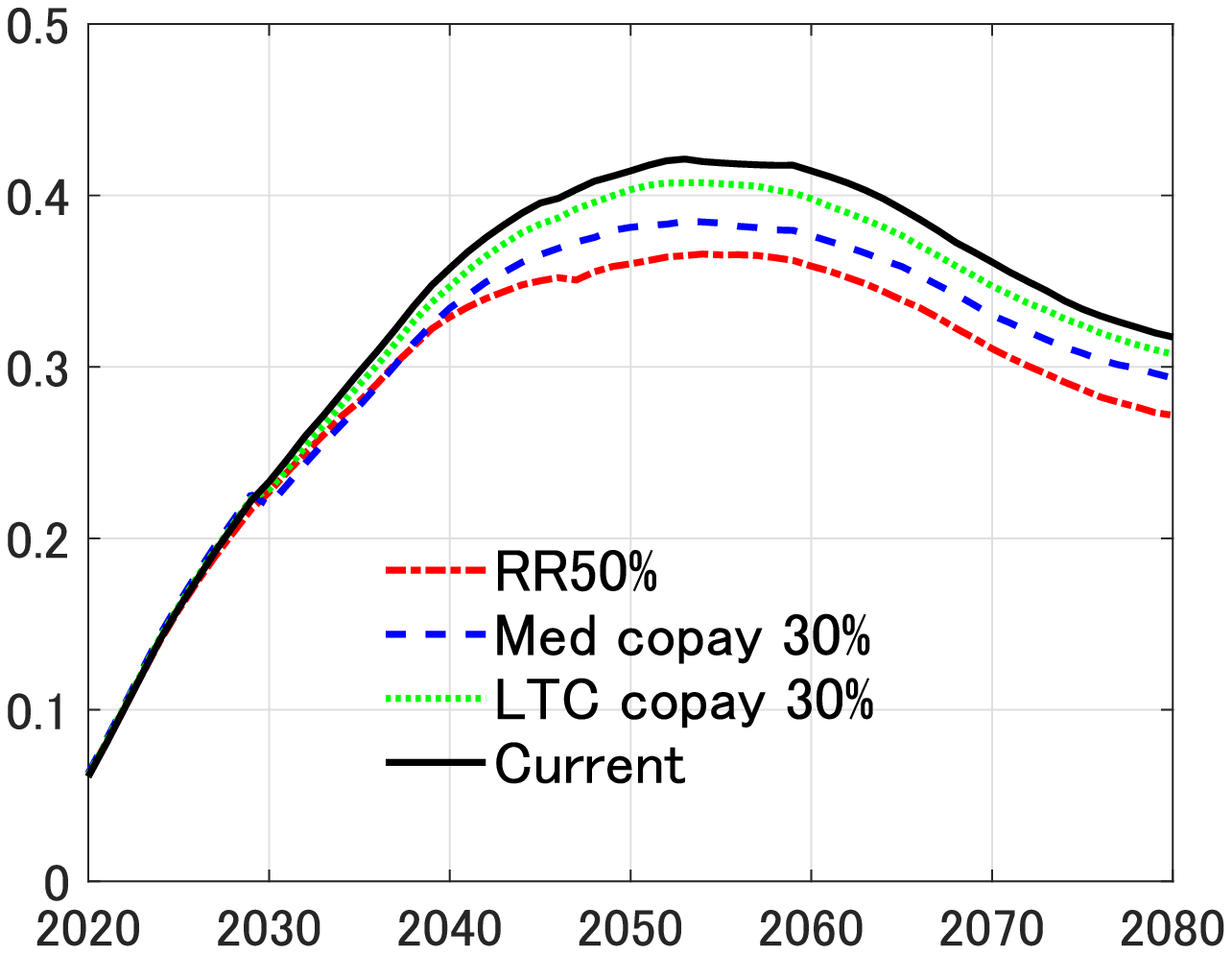}

\vspace{0.3\baselineskip}
\hspace{2em}
(a)~Baseline \hspace{11em}
(d)~Retirement extension
\end{center}
\end{minipage}
\caption{Transition path of consumption tax}
\label{fig:cons_tax_tran}
\vspace{0.5\baselineskip}
\end{figure}

%次に、消費税率のtransition pathに着目する。
%この結果は、図\ref{fig:cons_tax_tran}で示されており、現行政策維持の場合はどちらのシナリオでも、2050年半ばのピーク時点で消費税率が42～43\%程度まで上昇する。
%■Braun and Joines (2015) などの先行研究でも、人口成長率や政策シミュレーションの想定の違いはあるものの、おおむね整合的な傾向が示されている。
%\footnote{■Braun and Joines (2015)などにおける消費税率シミュレーションは、2060年代に45\%程度に達する結果が示されている。この理由は人口動態に関する仮定の違いだけでなく、我々のモデルでは特に無形資産を考慮しており資本の寄与が高くなっているために生じていると我々は考えられる。}。
Next, we turn to the consumption tax rates.
These transition paths are shown in Figure \ref{fig:cons_tax_tran}, where the consumption tax rates raises to around 42--43\% at the peak in the mid-2050s in both scenarios if the current policy is maintained.
Previous studies such as \citet{BJ2015JEDC}, have also shown generally consistent trends, albeit with differences in population growth rates and policy simulation assumptions.
\footnote{
The simulations of the consumption tax rates presented by \citet{BJ2015JEDC} and others show results that reach around 45\% in the 2060s. We think these are caused not only by differences in the assumptions about demographics, but also by the larger contribution of capital in our model, especially since it takes intangibles into account.}

%%%%%%%%%%%%%%%%%%%%%%%%%%%%%%%%%%%%%%%%%%%%
%%%%%%%%%%%%%%%%%%%%%%%%%%%%%%%%%%%%%%%%%%%%% 

%改革が消費税率に与える影響は、次の3点にまとめられる。
%第1に、どの改革も現行維持に比べて、ピーク時点の水準も低くなり、長期的にも税率が低下する。
%第2に、3つの改革の中では年金代替率引き下げの効果が最も大きく、どちらのシナリオでも2050年代半ばに36\％程度まで低下する。
%第3に、医療・介護支出の自己負担率引き上げの効果は、年金所得代替率引き下げの効果よりも小さい。加えて、介護支出の自己負担引き上げケースの効果はさらに効果が小さいく、どちらのシナリオでも、医療のケースでは2050年代半ばに38～39\％程度、介護のケースでは2050年代半ば41～42\％程度となる。
%\footnote{なお、■Braun and Joines (2015) では年金所得代替率引き下げよりも医療自己負担引き上げの方が消費税率を引き下げる効果が高いという結果が示されている。これには、社会保障改革に関する想定の違いが影響していると考えられる。同研究では、年金所得代替率の段階的な10\%引き下げ、70歳以上高齢者の医療支出自己負担率30\%への段階的な引き上げ、および社会保障以外の政府支出の段階的な10\％引き下げが想定されている。またsection \ref{sec:CALIB}で述べたように、我々のモデルでは労働時間が賃金に対して非弾力的な設定としているため、労働供給の反応の差が結果に違いをもたらしている可能性がある。}
The impact of the reforms on the consumption tax rate can be summarized as the following three points.
First, all reforms would also reduce the peak level and lower the tax rate in the long-run compared with the current policy.
Second, among the three reforms, the effect of reducing the pension replacement rate would be the largest, reducing the tax rates to about 36\% in the mid-2050s in both scenarios.
Third, the effect of rising the copayment rates for medical and LTC expenditures is smaller than the effect of reducing the pension income replacement rate.
In addition, the effect in the case of increasing the copayments for LTC expenditures is even smaller, falling to about 38--39\% in the mid-2050s in the case of health care and to about 41--42 \% in the mid-2050s in the case of LTC expenditures in both scenarios.
\footnote{
\citet{BJ2015JEDC} show that rising medical copayments is more effective in lowering the consumption tax rate than reducing the pension income replacement rate. This could reflect differences in assumptions about social security reform.
They suppose a gradual 10\% reduction in the pension income replacement rate, a gradual increase in the medical care copayment rate for the elderly over age 70 to 30\%, and a gradual 10\% reduction in government spending outside of social security.
Also, as noted in section \ref{sec:CALIB}, our model is that hours worked are set inelastic for wages, so differences in the labor supply response may lead to different results.}

\subsection{Welfare analysis}
%\subsubsection{Pension reform}
%次に、現状政策（current policy）のケースを基準に、2つのシナリオのもとで社会保障改革を実施した場合の人々の厚生への影響を、現在の勤労世代、引退世代と将来世代に区別してコホートごとに記述する。
%さらに、性別（gender）と雇用形態（employment type）で区分した4タイプの家計の異質性に着目した分析を行う。
Using the no-reform case as a benchmark, we assess the impact on the welfare of the three social security reforms for all generations, distinguishing the current working, the current retired, and the future generations in the two scenarios.  
We also consider the heterogeneity of the four types of households, distinguished by gender and employment type.

\begin{figure}[tp]
 \centering

\vspace{0.2\baselineskip}
\noindent\begin{minipage}[t]{1\columnwidth}%
\begin{center}

\hspace{2em}
\includegraphics[width=76.61mm]{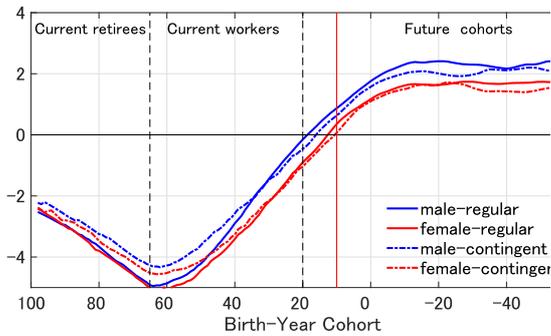} \hspace{0.5em}
\includegraphics[width=76.61mm]{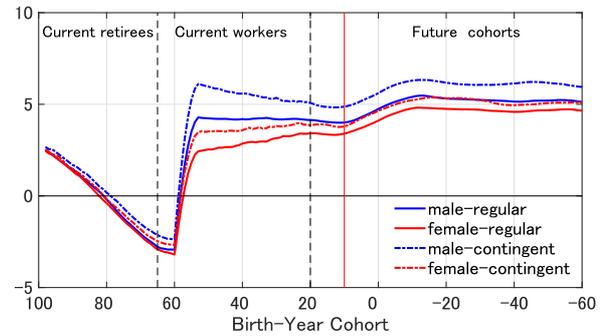}

\vspace{0.2\baselineskip}
\hspace{1em}
(a)~Baseline \hspace{12em}
(d)~Retirement extension
\end{center}

\end{minipage}

\caption{Welfare change due to replacement rate reduction}
\label{fig:welfare_RR}
%\vspace{0.5\baselineskip}
\begin{footnotesize}
\begin{itemize}
    \item[Note:] The vertical red line drawn from 10 on the x-axis indicates the year in which the medical and LTC reforms will take place. The figure shows the percentage change in welfare relative to the current policy.
\end{itemize}
\end{footnotesize}

\end{figure}

%第1に、年金所得代替率引き下げの影響を見る。
%この改革では、どちらのシナリオにおいて、現時点で定年年齢前後の人々の厚生が大きく下落している（図\ref{fig:welfare_RR}）。
%この年齢層の人々は、既に年金保険料の支払いの多くを済ませている一方で、給付が引き下げられていく時期に年金受給が開始になることが影響していると考えられる。
%年金改革では、移行期に当たる人々の損失に注意が必要であるが、そのことが我々の分析結果でも示されている。
%ベースライン・シナリオでは現在の引退世代、勤労世代のほぼすべての年代で厚生が低下している一方、定年延長シナリオでは、現役世代の厚生が2〜5\%程度改善している。
%また、年金所得代替率の引き下げでは、男女ともにフルタイマーの厚生がパートタイマーよりも大きく低下する傾向があることもわかる。
%これは、section \ref{sec:CALIB}でも述べたように、相対的に所得の高いフルタイマーは、年金保険料を多く支払っているにもかかわらず給付が減少することが要因だと考えられる。
%\paragraph{Pension reform}
First, let us deal with the reform involving a reduction in the pension income replacement rate. This reform results in a significant fall in the welfare of individuals around retirement age in the present in both scenarios (Figure \ref{fig:welfare_RR} Panel (a) and (b)).
We consider the impact to be that individuals in this age group, who have already paid most of their pension contributions, will begin to receive pension benefits when the benefit is reducing.
In the baseline scenario, welfare declines for almost all ages of the current retired and working generations, although in the extended retirement scenario, each type of agent's welfare improves by about 2--5\% for the working generation.
This pension reform also causes a more significant decline in benefits for regular workers than for contingent workers for both the male and the female group.
As discussed in section \ref{sec:CALIB}, we deduce that the reason for this is that regular workers, who have relatively higher incomes, pay more in pension contributions, despite receiving reduced benefits.

%%%%%%%%%%%%%%%%%%%%%%%%%%%%%%%%%%%%%%%%%%%%
%%%%%%%%%%%%%%%%%%%%%%%%%%%%%%%%%%%%%%%%%%%%% 

\begin{figure}[tb]
 \centering

\noindent\begin{minipage}[t]{1\columnwidth}%
\begin{center}

\vspace{0.2\baselineskip}
\hspace{2em}
\includegraphics[width=76.61mm]{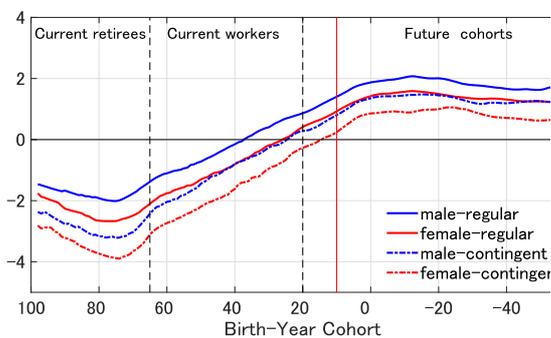} \hspace{0.5em}
\includegraphics[width=76.61mm]{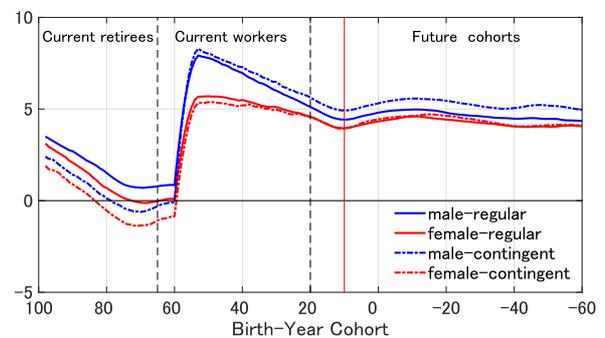}

\vspace{0.2\baselineskip}
\hspace{1em}
(a)~Baseline \hspace{12em}
(d)~Retirement extension

\end{center}
\end{minipage}

 \caption{Welfare change due to medical copayment up to 30\%}
 \label{fig:welfare_MED}

\begin{footnotesize}
\begin{itemize}
    \item[Note:] The vertical red line drawn from 10 on the x-axis indicates the year in which the medical and LTC reforms will take place. The figure shows the percentage change in welfare relative to the current policy.
\end{itemize}
\end{footnotesize}

\end{figure}

%\subsubsection{Medical insurance reform}
%第2に、医療保険改革が厚生に及ぼす効果を確認する。
%図\ref{fig:welfare_MED}は、ベースライン・シナリオにおいて年金改革と同様の傾向を示しているが、特に現在の引退世代の厚生が低下が顕著である。
%また、この改革では男女ともにパートタイマーの方が厚生の低下が大きく、女性のパートタイマーは高々4\%の低下である。
%定年延長シナリオの場合は、年金改革と同じく現在の勤労世代の厚生が改善している。
%\paragraph{Medical insurance reform}
Next, we move on to the reform rising the copayment rate in the medical insurance.
Figure \ref{fig:welfare_MED} Panel (a) shows a similar trend to the pension reform in the baseline scenario. 
In particular, there is a significant decline in the welfare of the current retirees.
The reform also shows a substantial decrease in welfare for both male and female contingent workers. 
For instance, the welfare of female contingent workers declines by at most 4\%.
In the extended retirement scenario, similar to the pension reform, the welfare of the current working-age population improves (Figure \ref{fig:welfare_MED} Panel (b)).

%%%%%%%%%%%%%%%%%%%%%%%%%%%%%%%%%%%%%%%%%%%%
%%%%%%%%%%%%%%%%%%%%%%%%%%%%%%%%%%%%%%%%%%%%% 

%\subsubsection{LTC insurance reform}
%第3に、介護保険改革が厚生に及ぼす影響について説明する。
%ベースライン・シナリオにおいて、図\ref{fig:welfare_LTC}は、医療のケースと比べてより高齢な層の厚生が低下していることを示している。
%これは、section \ref{sec:CALIB}で述べたように、医療と比べて介護支出の方が90歳以上の1人当たり支出が高いことと整合的である。
%定年延長シナリオでは、医療改革と同様の傾向が示されている。

\begin{figure}[tp]

\noindent\begin{minipage}[t]{1\columnwidth}%
\begin{center}

%\vspace{0.2\baselineskip}
\hspace{2em}
\includegraphics[width=76.61mm]{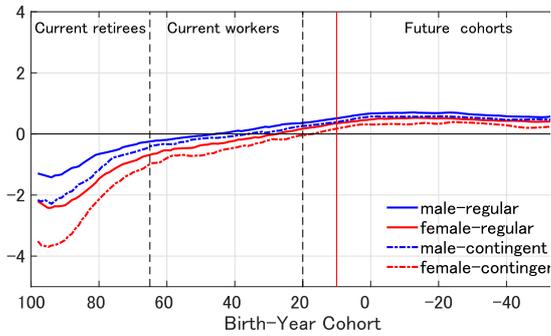} \hspace{0.5em}
\includegraphics[width=76.61mm]{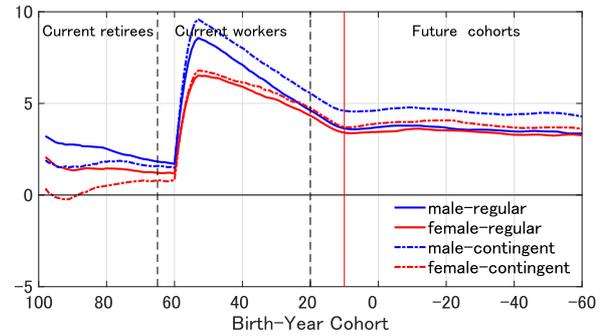}

\vspace{0.2\baselineskip}
\hspace{1em}
(a)~Baseline \hspace{12em}
(d)~Retirement extension

\end{center}
\end{minipage}

 \centering
 \caption{Welfare change due to LTC copayment up to 30\%}
\label{fig:welfare_LTC}

\begin{footnotesize}
\begin{itemize}
    \item[Note:] The vertical red line drawn from 10 on the x-axis indicates the year in which the medical and LTC reforms will take place. The figure shows the percentage change in welfare relative to the current policy.
\end{itemize}
\end{footnotesize}

\end{figure}

%\paragraph{LTC insurance reform}
Finally, we describe the reform increasing the copayment rate of LTC insurance. In the baseline scenario, Figure \ref{fig:welfare_LTC} Panel (a) shows a decline in welfare for the current older age groups compared with the medical care case.
This result is consistent with the LTC expenditures being higher per capita expenditures for those aged 90 and older compared with health care, as discussed in section \ref{sec:CALIB}.
The extended retirement age scenario shows a similar trend to the medical insurance reform (Figure \ref{fig:welfare_LTC} Panel (b)).

%%%%%%%%%%%%%%%%%%%%%%%%%%%%%%%%%%%%%%%%%%%%
%%%%%%%%%%%%%%%%%%%%%%%%%%%%%%%%%%%%%%%%%%%%% 

%3つの改革に基づく厚生分析の結果は、表〓にまとめられている。
%ここでは、Totalと3つの世代――Retire, Worker, Future――ごとに、それぞれの厚生変化の平均値を示している。
%我々の分析では、ベースラインシナリオにおいては3つの改革はいずれもfuture generationsの厚生を増加させるが、current retired and working generationsの厚生を低下させることがわかった。特に、医療・介護改革におけるfemale contingent workersの平均的な厚生の低下は2.3—3.4\%と相対的に大きい。
%それに対して、定年延長シナリオにおいてはcurrent generationsの厚生低下が軽減されていることが明らかとなった。
The size of the changes in welfare based on the three reforms ("RR50", "Med10", "LTC10") compared to the current system is summarized in Table \ref{tab:welfaretable}. 
The item for "Total" is the size of the change in welfare aggregated across the three generations over the entire period, while the items for "Retire", "Worker", and "Future" are those for each of the generations.
Positive values of item "Future" in the table indicates that all three reforms increase the welfare of all four agents of future generations, while negative values of "Retiree" and "Worker" show they decline the welfare of all four agents of the current retired and working generations in the baseline scenario. 
In particular, the average welfare drop of female contingent workers in the medical and LTC reforms is significant, amounting to minus 2.3--3.4 percent.
However, these reforms in the extension of the retirement age scenario eliminate the decrease in welfare of current generations. Indeed, there are all positive values for all agents except a few of item "Retire" in the columns of "Extension of Retirement Age" in the table.

\begin{table}[t]%[hbtp]
\begin{footnotesize}
\caption{Summary of welfare analysis}
\label{tab:welfaretable}

\begin{center}
\begin{tabular}{ll|rrrr|rrrr}
\hline
\textbf{}                                    & \cellcolor[HTML]{FFFFFF}\textbf{} & \multicolumn{4}{c|}{\cellcolor[HTML]{FFFFFF}\textbf{1. Baseline}}                                                                                                                                                  & \multicolumn{4}{c}{\cellcolor[HTML]{FFFFFF}\textbf{2. Extension of Retirement Age}}                                                                                                                               \\
\rowcolor[HTML]{FFFFFF} 
\multicolumn{1}{c}{\cellcolor[HTML]{FFFFFF}} &                                   & \multicolumn{1}{c}{\cellcolor[HTML]{FFFFFF}Total} & \multicolumn{1}{c}{\cellcolor[HTML]{FFFFFF}Retire} & \multicolumn{1}{c}{\cellcolor[HTML]{FFFFFF}Worker} & \multicolumn{1}{c|}{\cellcolor[HTML]{FFFFFF}Future} & \multicolumn{1}{c}{\cellcolor[HTML]{FFFFFF}Total} & \multicolumn{1}{c}{\cellcolor[HTML]{FFFFFF}Retire} & \multicolumn{1}{c}{\cellcolor[HTML]{FFFFFF}Worker} & \multicolumn{1}{c}{\cellcolor[HTML]{FFFFFF}Future} \\ \hline
\rowcolor[HTML]{FFFFFF} 
RR50  &  male-regular  & $-$0.325  & $-$3.666  & $-$2.857  & 2.008  & 3.555  & $-$0.129  & 3.259  & 4.993  \\
\rowcolor[HTML]{FFFFFF} 
  & female-regular  & $-$0.787  & $-$3.707  & $-$3.344  & 1.408  & 2.921  & $-$0.234  & 2.090  & 4.414  \\
\rowcolor[HTML]{FFFFFF} 
  & male-contingent  & $-$0.304  & $-$3.195  & $-$2.601  & 1.762  & 4.434  & 0.287  & 4.506  & 5.867  \\
\rowcolor[HTML]{FFFFFF} 
  & female-contingent  & $-$0.751  & $-$3.408  & $-$3.137  & 1.273  & 3.420  & 0.000  & 2.830  & 4.897  \\ \hline
\rowcolor[HTML]{FFFFFF} 
Med10    & male-regular  & 0.554  & $-$1.741  & $-$0.184  & 1.700  & 4.355  & 1.776  & 5.910  & 4.561  \\
\rowcolor[HTML]{FFFFFF} 
  & female-regular  & 0.037  & $-$2.348  & $-$0.807  & 1.263  & 3.665  & 1.081  & 4.600  & 4.154  \\
\rowcolor[HTML]{FFFFFF} 
  & male-contingent  & $-$0.137  & $-$2.852  & $-$0.930  & 1.184  & 4.485  & 0.538  & 6.105  & 5.144  \\
\rowcolor[HTML]{FFFFFF} 
  & female-contingent  & $-$0.688  & $-$3.426  & $-$1.631  & 0.709  & 3.402  & $-$0.123  & 4.283  & 4.248  \\ \hline
\rowcolor[HTML]{FFFFFF} 
LTC10    & male-regular  & 0.192  & $-$0.799  & 0.075  & 0.597  & 3.992  & 2.455  & 6.114  & 3.571  \\
\rowcolor[HTML]{FFFFFF} 
  & female-regular  & $-$0.131  & $-$1.589  & $-$0.227  & 0.428  & 3.421  & 1.429  & 5.058  & 3.382  \\
\rowcolor[HTML]{FFFFFF} 
  & male-contingent  & $-$0.001  & $-$1.308  & $-$0.061  & 0.489  & 4.555  & 1.680  & 6.899  & 4.506  \\
\rowcolor[HTML]{FFFFFF} 
  & female-contingent  & $-$0.426  & $-$2.358  & $-$0.463  & 0.274  & 3.501  & 0.386  & 5.269  & 3.798                                   \\ \hline
\end{tabular}
\end{center}
\begin{itemize}
    \item[Note:] The item for "Total" is the size of the change in welfare aggregated across the three generations over the entire period, while the items for "Retiree", "Worker", and "Future" are those for each of the generations.
\end{itemize}

\end{footnotesize}
\end{table}
 %厚生分析のまとめ表

\subsection{Discussion}%5.4
%われわれは、シミュレーションの結果の中で次の3点に着目する。
%第１に、高齢化とともに実質GDP成長率は21世紀半ばまで減少するが、その減少への寄与は労働力の減少よりも資本ストックの減少が大きいことがわかった（図9）。
%第2に、われわれの厚生分析では、ベースラインのシミュレーションにおいて、社会保障改革は現在世代の厚生を引き下げるインパクトをもたらすことがわかった。
%ただし、その影響の大きさは家計のタイプによって異なる。
%年金改革はフルタイマーの厚生を4\%程度引き下げる一方、パートタイマーの厚生は1\%程度引き下げる。つまり、フルタイマーの方が負の影響が大きい。
%医療・介護改革は、女性やパートタイマーの厚生を2--3\%程度引き下げる一方、男性のフルタイマーの厚生は1\%程度に低下する。つまり、女性やパートタイマーの方が負の影響が大きい。
%第3に、社会保障改革と同時に定年延長を行うことで、現在世代の厚生の低下を3--4\%程度緩和することができる。
Here, we pay more attention to the following three aspects based on our simulation results.
First, we find that real GDP growth would decline with aging until the mid-21st century and that most of the decline appears to be attributable to the fall in the capital stock rather than to the decline in the labor force (Figure \ref{fig:GDPtran_contribution}). 
Second, our welfare analysis shows that  social security reform has a low impact on the welfare of the current generation in our baseline  scenario.
However, the magnitude of the impact varies by household type.
Pension reform lowers the welfare of regular workers by about 4\%, while it lowers the welfare of contingent workers by about 1\%. In other words, the negative impact is more significant for regular workers.
Medical and LTC reform lowers the welfare of women and contingent workers by about 2 to 3\%, while it lowers the welfare of male regular workers by about 1\%. In other words, the negative impact is more significant for women and contingent workers.
Third, extending the retirement age at the same time as social security reform would mitigate the current generation's decline in welfare by about 3 to 4\%.

%これらの結果は、社会保障改革とともに定年延長も一緒に行うことが、現在世代の厚生低下の緩和のために重要であることを示唆している。
%また、われわれのシミュレーションは、社会保障改革は現在世代の厚生を引き下げること、加えて現在世代の中でも相対的に所得の低い層（女性、パートタイマー）により強いネガティブなインパクトを示している。
These counterfactual simulations suggest that extending the retirement age combined with social security reform is necessary to mitigate the decrease in the welfare of the current generation.
Our simulations also show that social security reform lowers the welfare of the current generations, with a more negative impact on the lower-income segments (women and contingent workers).

%それでは、なぜ、年金改革はフルタイマーの厚生を相対的に大きく引き下げる一方、医療・介護改革は、相対的に所得の低いタイプの家計の厚生を大きく引き下げるのだろうか。
%その要因の一つは、年金給付額が勤労期の自身の労働所得に依存していることである。
%相対的に所得の高いフルタイマーは、年金所得代替率が引き下げられたことによる給付額の減少が大きいため、厚生が低下したと考えられる。
%一方、われわれのシミュレーションでは、医療・介護支出は所得に依存せず、年齢と性別差に依存して発生すると仮定している。そのため、所得のより低いタイプの家計の厚生低下につながったと考えられる。
Why, then, does the pension reform reduce the welfare of \textit{regular workers} to a relatively large extent, while medical and LTC reform reduce the welfare of \textit{households belonging to lower income group} to a large extent?
One of the most significant reasons is that pension benefits depend on labor income during the working age.
Regular workers with relatively higher incomes are likely to experience a decline in welfare because of the significant decrease in benefits due to the reduction in the pension income replacement rate.
In our calculations, on the other hand, we suppose that medical and LTC expenditures are independent of income and arise depending on age and gender differences. 
Therefore, we suppose that the medical and LTC reforms led to a decline in welfare for households with lower income types.

%社会保障給付の削減のような老後の負担を引き上げる政策は、若年期の消費・貯蓄行動にも影響を及ぼす。特に、老年期の医療・介護支出が増えることが予想される場合は、若年期から貯蓄を増やすだろう（e.g. ■De Naldi et al 2010; Kopecky and Koreshkova 2014）。
%また、老後の健康リスクや所得リスクも、若年期の行動に影響を及ぼす。■Braun et al. (2017) では、老後の多様なリスクの存在を考慮すると、特に所得の低い家計への公的生活保障の拡大が厚生を改善することを指摘している。われわれのシミュレーションでは、社会保障改革は、相対的に所得が低く、年齢とともに所得が上昇しない家計の厚生を大きく引き下げている。われわれの結果は、この点では彼らの結果と整合的である。
%そのため、社会保障改革の実施においては、低所得層（女性、パートタイマー）への再分配による厚生低下の緩和が重要となる。
Health and income risks in old age affect behavior in younger age. 
\citet{Braun&KK2017} pointed out that, given the various risks in old age, increasing social security benefits by expanding means-tested social insurance improves welfare. Meanwhile, in our simulations, social security reform significantly lowers welfare for households with relatively low incomes and whose incomes do not increase with age. 
Our simulations could be consistent with theirs in this regard.
Therefore, when reforming social security, it is necessary to mitigate the decline in welfare by redistributing benefits to low-income groups such as females and contingent workers.

%ただし、本研究には次のような課題がある。
%われわれのモデルは、男性、女性、フルタイマー、パートタイマーのすべてのタイプについて単身の家計を想定している。
%しかし、■Doepke and Tertilt (2016) でまとめられているように、近年の研究では家計の中に結婚した男女、あるいは子どもなどが存在するモデルを用いて、さまざまなショックに対する家族間の保険機能を考慮した分析が行われている（e.g. ■Blundell_etal._2018_JPE; ■Fernández&Wong(2014EJ); ■Kitao and Mikoshiba, 2022）。
%例えば、女性が低所得であっても、彼女の配偶者が高所得である場合には、当然単身の女性とは意思決定が異なる。そのため、社会保障改革の影響を議論する際には家族の機能が重要な役割を果たす。
However, our study would have the following challenges.
Our model supposes single households for all types of gender and employment types.
However, as summarized in \citet{doepke&tertilt2016fams}, recent studies have used models with married couples or children in the household and have considered the insurance effect among family members against various shocks
\citep[e.g.,][]{blundellChildrenTimeAllocation2018, fernandezDivorceRiskWages2014, KSMM2022FLFP}.
For example, if a woman has a low income, but her spouse has a high income, her decision-making will naturally differ from that of a single woman. Therefore, we suppose that family functioning plays an essential role in considering social security reform's impact in more detail.

%%%%%%%%%%%%%%%%%%%%%%%%%%%%%%%%%%%%%%%%%%%%%%%%%%%%%%%%%%%%%%%%%%%%%%%%%%%%%%%%%%%%
\section{Conclusion}\label{sec:CONCL}

This paper quantitatively explores the impact of pension, medical, and LTC insurance reforms on households' welfare in Japan, where the population is aging remarkably, through policy simulations using an OLG model with four types of households distinguished by gender and employment type.

Our simulations show that all three options of social security reforms raise the welfare of future generations, in contrast to having a significantly negative impact on low-income groups, females and contingent workers, in the current generations.
We also find that these reforms affect females negatively.
However, combining the extension of the retirement age with social security reforms mitigates the negative impact on the current generations' welfare by 3 to 4\%.
The result suggests that the negative impact of social security reforms on welfare could be reduced by paying enough attention to the impact on females and contingent workers, who are relatively vulnerable groups.

However, there are the following challenges in our study.
Firstly, since future medical and LTC expenditures depend on uncertainty, such as individual health status, there are many implications of focusing on future uncertainty, especially in the context of the retirement saving puzzle, for which the literature has recently accumulated  \citep[e.g.,][]{DeNardietal2010, KKLTC2014}. 
%%(e.g., ■ DeNaldi et al 2010, ■ Kopecky and Koreshkova 2014).
%%(natbib e.g.の出し方 https://tex.stackexchange.com/questions/166410/how-can-i-include-e-g-in-citep)
In contrast, our model is that households behave deterministically.
The next challenge is that our model omits the effect of risk sharing on future uncertain spending among married couples.
As mentioned in the previous section, some recent research focuses on such family effects, and \citet{Braun&KK2017} quantifies the risk of becoming single after retirement.

Analyzing such effects more rigorously requires microdata on such ad household consumption, income, medical, and LTC expenditures.
Our analysis, however, is based on aggregate data.
We may be able to provide more practical and meaningful policy implications for social security reforms while mitigating negative impacts as much as possible by overcoming these challenges.
We want to make these challenges our next research subject.

%この研究では、高齢化が顕著に進む日本において、年金・医療・介護制度改革の実施が人々の厚生に及ぼす影響を、性別と雇用形態によって4つのタイプにエージェントが存在する世代重複モデルを用いた政策シミュレーションを通じて定量的に分析した。

%その結果、いずれの社会保障改革も将来世代の厚生を一貫して引き上げる一方で、特に現役世代における相対的な低所得者（女性、パートタイマー）への負の影響が大きいことがわかった。
%また女性は、全ての社会保障改革において、一貫して大きな負の影響を受けるという結果が得られた。
%加えて、我々は定年延長を同時に行うことで、特に現役世代への負の影響が緩和され、厚生が2～5\%改善することも見出した。
%我々の政策シミュレーションは、将来的に社会保障改革を行う際には、相対的に社会的弱者である女性、パートタイマーへの影響に十分に配慮することで、社会厚生の低下を抑えられる可能性を示唆している。

%ただし、我々の研究には次に述べる課題がある。
%まず、将来の医療・介護支出は個人の健康状態などの不確実性な要因に左右されるため、将来の不確実性に着目することで得られる含意は多く、特にretirement saving puzzleの文脈で多くの研究が蓄積されている（例：■DeNaldi et al 2010、■Kopecky and Koreshkova 2014）。しかし、我々のモデルは確定的である。
%また、我々のモデルでは、結婚を通じた将来支出のリスクシェアリングなどの影響も捨象している。たとえば■Braun, Kopecky and Koreshkova(2017)では、引退期にシングルとなることのリスクを定量的に分析している。このような影響をより厳密に分析するためには、家計の消費、所得、医療・介護等に関するミクロデータを用いる必要がある。しかし、我々の分析は集計データに基づいている。
%これらの課題を改善することで、負の影響をできるだけ緩和しつつ社会保障改革を実施するための、より現実的(realistic)で重要な示唆が得られるだろう。我々はこれを次のリサーチのテーマとしたい。

%%%%%%%%%%%%%%%%%%%%%%%%%%%

\vspace{1\baselineskip}
%%Refereces%%%%%%%%%%%%%%%%%%%%%%%%%%%%%%%%%
\renewcommand{\baselinestretch}{0.7} \normalsize
\bibliographystyle{econ} %econ.bst by Shiro Takeda

\begin{small}
\bibliography{olg_dtc}
\end{small}
%%%%%%%%%%%%%%%%%%%%%%%%%%%%%%%%%%

%\appendix
%\input{References_2.tex}

%\newpage

%\section*{Tables}

%\input{Tables1-4_2.tex}

%\newpage

%\input{Figures_2.tex}

%%%%%%%%%%%%%%%%%%%%%%%%%%%%%%%%%%%%%
\end{document}